# X-ray view of dissipative warm corona in active galactic nuclei

B. Palit⋆[1], A. Różańska[1], P. O. Petrucci[2], D. Gronkiewicz[1], S. Barnier[6], S. Bianchi[4], D. R. Ballantyne[5], V. E. Gianolli[2,4], R. Middei[7,8], R. Belmont[3], and F. Ursini[2]

[1] Nicolaus Copernicus Astronomical Center, Polish Academy of Sciences, Bartycka 18, 00-716 Warsaw, Poland
[2] Université de Grenoble Alpes, IPAG, F-38000 Grenoble, France
[3] Université Paris Cité, Université Paris-Saclay, CEA, CNRS, AIM, F-91191, Gif-sur-Yvette, France
[4] Dipartimento di Matematica e Fisica, Università degli Studi Roma Tre, via della Vasca Navale 84, I-00146 Roma, Italy
[5] Center for Relativistic Astrophysics, School of Physics, Georgia Institute of Technology, 837 State Street, Atlanta, GA 30332-0430, USA
[6] Theoretical Astrophysics, Department of Earth and Space Science, Graduate School of Science, Osaka University, Toyonaka, Osaka 560-0043, Japan
[7] Space Science Data Center, Agenzia Spaziale Italiana, Via del Politecnico snc, 00133 Roma, Italy
[8] INAF Osservatorio Astronomico di Roma, Via Frascati 33, 00078 Monte Porzio Catone (RM), Italy

**ABSTRACT**

*Context.* Detections of the X-ray spectra of active galactic nuclei (AGNs) typically reveal a noticeable excess of soft X-rays beyond the extrapolation of the power-law trend observed between 2 and 10 keV. However, the cause of this surplus remains unclear. In the scenario of soft Comptonization, observations suggest a warm corona temperature, ranging from 0.1 to 1 keV, and an optical depth of approximately 10 to 30. Furthermore, according to radiative constraints derived from spectral analyses employing Comptonization models, it has been suggested that most of the accretion power is released within the warm corona. At the same time, the disk beneath it is largely non-dissipative, mainly emitting the reprocessed radiation from the corona.

*Aims.* We tested the dissipative warm corona model using the radiative transfer code `TITAN-NOAR` on a sample of 82 *XMM-Newton* EPIC-pn observations of 21 AGNs. Based on a spectral modeling of the X-ray data, we aim to estimate the total amount of internal heating inside the warm corona on top of the accretion disk.

*Methods.* By modeling the 0.3–10 keV EPIC-pn spectra with the `TITAN-NOAR` model component, we estimated the internal heating and optical depth of the warm corona and checked their correlations with such global parameters as the hot corona spectral index, black hole mass, and accretion rate. From the model normalization, we computed the radial extent of the warm corona on top of the cold accretion disk.

*Results.* Our model infers the presence of dissipative warm corona, with optical depths distributed across the range of $\sim$ 6–30 and a total internal heating in the range of $\sim 1\text{–}29 \times 10^{-23}$ erg s$^{-1}$ cm$^3$. We do not detect any variation between these properties and global properties, such as the black hole mass and accretion rate. The extent of the warm corona is spread across a broad range, from 7–408 gravitational radii, and we find that the warm corona is more extended for higher accretion rates.

*Conclusions.* Soft excess emission is ubiquitous across a wide mass range and accretion rate in AGNs. We confirm that the warm corona responsible for producing the soft X-ray excess is highly dissipative with greater optical depths being associated with lower internal heating and vice versa. The cold standard accretion disk regulates the extent of the warm corona.

## 1. Introduction

Despite major advancements in understanding the X-ray spectral features of active galactic nuclei (AGNs), the origin of soft X-ray excess (Pravdo et al. 1981) remains puzzling. Typically, the X-ray spectra of Seyfert 1 galaxies are dominated by primary emission in the form of hard X-ray power law above 2 keV, originating in the hot, optically thin plasma close to the central black hole. When the hard power law is extrapolated below 2 keV, a smooth excess rises above it, known as the soft X-ray excess.

Generally well described by a Comptonized emission, the electron temperature of "soft X-ray excess" peaks at $\sim$ 0.1 - 1 keV for AGNs covering a wide range of accretion rates, black hole masses, and activity type (Czerny et al. 2003; Done et al. 2012). Such a constant trend with respect to temperature points towards a similar origin of emission for different types of AGNs, which may be connected with reprocessing of the X-ray emission. This resulted in associating either blurred reflection (Fabian et al. 2004; Crummy et al. 2006) or blurred ionized wind absorption (Gierliński & Done 2004) as probable origins of the soft X-ray excess. In the case of the blurred ionized disk reflection model, intrinsic hard X-rays are focused on the accretion disk producing a reflection continuum with a dense forest of emission lines which are then relativistically broadened due to proximity to the supermassive black hole (SMBH). While it nicely reproduces the smooth shape of soft X-ray excess and also demonstrates a physical connection with spectral turnover at $\sim$ 30 keV (known as the Compton reflection hump), it is expected that soft X-ray excess strength must be correlated with the strength of reflection. However, the opposite relation was observed by Boissay et al.

*Send offprint requests to*: B. Palit
⋆ E-mail: bpalit@camk.edu.pl (BP)

(2016), who determined that reflection factor anti-correlates with the strength of the soft X-ray excess in the sample of about 102 sources. In addition, the high spin of the black hole required and high disk density ($\geq 10^{18}$ cm$^{-3}$) inferred from the model raise questions about its feasibility. The same conclusion was reached for the smeared absorption model, which requires a very high speed of an ionized wind, up to around $0.9c$, to provide sufficient relativistic blurring to fit the spectra correctly (Schurch & Done 2008).

One plausible way to connect the soft X-ray excess with atomic data is the model of Compton reflection from a pure hydrogen atmosphere found by Madej & Różańska (2000), where authors demonstrated that Compton scattering can shift high energy photons toward lower energies during reflection from fully ionized matter consisting of hydrogen only. The soft X-ray excess arises when the lack of heavy elements prevents the absorption of soft photons re-emitted in the process of Compton down-scattering. Nevertheless, this result was not explored further due to its complexity and the X-ray data invoked simpler phenomenological solutions. To match the hard energy tail of the soft X-ray excess, an additional Comptonization component can be used when fitting the data (Magdziarz et al. 1998; Mehdipour et al. 2011; Jin et al. 2012; Petrucci et al. 2013, 2018; Porquet et al. 2018; Tripathi et al. 2021, and references therein). This led to the argument that soft X-ray excess could also arise from a separate Comptonizing medium, where a warm (electron temperature $kT_e \sim 0.1$ - 1 keV), optically thick (optical depth $\tau$ > 1) corona (distinct from the hot optically thin corona) is responsible for Compton up-scattering of seed photons from the disk (optical/UV energy range) and producing the characteristic shape of soft X-ray excess (Magdziarz et al. 1998; Done et al. 2012; Petrucci et al. 2013). This additional Comptonized layer, referred to as the warm corona, may be considered as a radial zone separate from standard disk (Done et al. 2012; Kubota & Done 2018) or a warm optically thick layer on top of the standard disk (Janiuk et al. 2001; Różańska et al. 2015; Gronkiewicz et al. 2023). Correlation found between UV/X-ray strongly support this interpretation (Mehdipour et al. 2011; Noda et al. 2011, 2013; Petrucci et al. 2013; Gliozzi & Williams 2020). Most of the models fitted to observations are phenomenological and lack physical grounds for the origin of such a warm layer.

Application of the soft Comptonization model on extensive X-ray observations yielded the presence of an optically thick layer of depth 10-30 and electron temperature $\sim$ 0.1-1.0 keV (Magdziarz et al. 1998; Page et al. 2004; Mehdipour et al. 2011; Jin et al. 2012; Petrucci et al. 2013; Matt et al. 2014; Mehdipour et al. 2015; Porquet et al. 2018; Ursini et al. 2018; Middei et al. 2018, 2019). By modeling the energy balance between the warm corona and cold disk, it was proven by Różańska et al. (2015) that such warm corona, cooled by Comptonization, has to be additionally heated by some internal process, most probably mechanical heating, to stay in hydrostatic equilibrium with a cold accretion disk. Furthermore, it was suggested that the existence of magnetic pressure or mass outflow was required to stabilize a warm corona of optical depth larger than 5. The above consideration did not specify any particular heating, it only showed that corona must dissipate energy to be constantly visible, as observed in Mrk 509 (Petrucci et al. 2013).

On the other hand, it was pointed out by García et al. (2019) that the emergent spectra from optically thick layer should carry strong signatures of absorption lines in soft X-ray spectrum, which contradicts the smooth shape of soft X-ray excess. However, when theoretical models of warm corona emission were computed including additional mechanical heating of plasma, the modeled spectra appeared featureless, which is in agreement with observations (Petrucci et al. 2020; Xiang et al. 2022). It is most likely that the excess heating raises the ionization state of matter in the optically thick warm corona, in turn reducing the photo-electric opacity. This smoothens the absorption features. Thus, a new question arises regarding the physical justification of the energy dissipation in the warm corona and whether can we estimate the amount of warm corona heating from observations.

Recently, a follow-up of optical/UV continuum emission was seen to track the changes in soft X-ray excess, suggesting a link between intrinsic disk emission and its interaction with warm corona producing the soft X-ray excess (Mehdipour et al. 2023). Such findings confirm our model of the dissipative flow, where both vertical layers (the warm corona and cold disk) are heated by magneto-rotational instability (MRI) and radiatively coupled (Gronkiewicz et al. 2023). Such additional heating ensures that scattering dominates over photoelectric absorption, seen as the smoothing sharp features in the emergent spectra. However, the above model only shows that warm corona and cold disk layers can coexist in equilibrium, both self-consistently heated by MRI according to the scheme proposed by Begelman et al. (2015), where the transition between layers is justified by stating global boundary conditions. Current codes cannot self-consistently produce dissipative warm corona coupled with an accretion disk. They only include additional heating in the energy balance equation of the warm layer cooled by the Comptonization of soft photons. To produce the spectra for data fitting, advanced radiative transfer codes should be used. Such spectral models recently became available with the `ReXcor` model (Xiang et al. 2022) and by the `TITAN-NOAR` code (Petrucci et al. 2020), where the former model was tested with real data (Porquet et al. 2024; Ballantyne et al. 2024). However, the former was tested only very recently and for only one source HE1029-1401 (Vaia et al. 2024).

In this work, we test the recent warm corona emission model computed by radiative transfer code `TITAN-NOAR` (Petrucci et al. 2020), on a sample of 21 AGNs observed with the *XMM-Newton* satellite over last two decades. In total, 82 observations have been analyzed in this paper, thus allowing for observational constraints to be placed on the amount of internal heating required by warm corona to sustain hydrostatic equilibrium with a cold accretion disk. The main assumptions of a warm, dissipative corona in our model together with a comparison with `ReXcor` model are present in Sect. 2, including the approach of numerical computations of the soft X-ray excess emission. In Sect. 3, we outline our AGN sample. In Sect. 4, we elaborate on the total model considered in the data fitting process. An interpretation of our results is given in Sect. 5, followed by a discussion and conclusions in Sects. 6 and 7, respectively.

## 2. General model assumptions

Figure 1 presents a toy model describing different emission components contributing to the total X-ray spectra. We have considered only radio-quiet sources, with negligible jet emission. Therefore, all observed X-ray emission originates from the inner accretion flow, which are potentially multiphase in nature. The inner accretion flow geometry is often described using a stratified model consisting of two separate regions of plasma: the one nearest to a black hole is the hot corona (shown in red) and the one next to the hot corona is the second zone, lying on the top of a cold accretion disk and known as the warm corona (in blue), thereby forming the so-called sandwich geometry (Haardt & Maraschi 1993). The warm corona is optically thick and is cooled in the process of up-scattering the soft-seed photons from the standard disk. It is believed to produce the soft X-ray excess which is seen in the energy range of 0.1 - 2 keV. The transition radius between single-phase hot flow and two-phase warm corona coupled with cold disk is not known. To make its measurement possible, we need to specify the physical process responsible for such transition. Past research has been focused on the disk evaporation model as a smooth change between cold accretion flow and hot corona (Różańska & Czerny 2000a,b), but this model still requires further theoretical exploration.

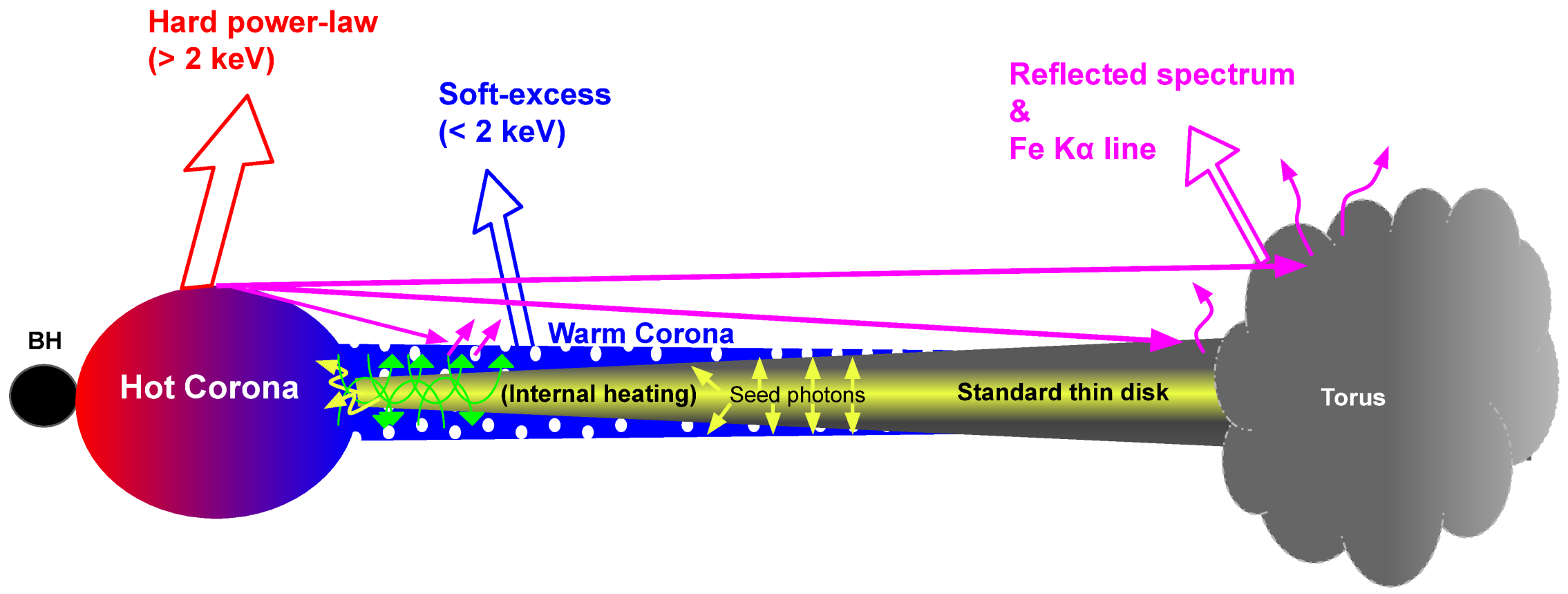

Fig. 1: Different emitting components considered in our data analysis of AGNs hosting a black hole in its center. The inner hot corona (red) is responsible for continuum emission above 2 keV in the form of a power-law. The hard X-ray radiation illuminates each part of the inner disk and the reflection is self-consistently computed by our code, `TITAN-NOAR`. The reflected emission (magenta arrows) can originate from the outer disk (yellow) as well as from the distant torus (grey). The two-phase slab of warm corona (blue layer zone) and cold disk (yellow) is responsible for producing soft X-ray excess emission by up-scattering optical/UV seed photons (shown as yellow arrows) from the disk. Additional internal heating in this region assumed in our model is marked with curved green arrows.

This work comprises an attempt to study the warm corona in terms of the amount of energy dissipated in this layer which is situated on the top of a cold disk. The outgoing spectrum is calculated with the radiative transfer code `TITAN` (Dumont et al. 2003), which fully solves the transfer of radiation through the stratified matter with heavy elements by taking into account additional mechanical heating in energy balance to justify temperature equilibrium. Simultaneously, an ionization balance in non-local thermal equilibrium (non-LTE) regime and Compton scattering are fully taken into account with the Monte Carlo code `NOAR` (Dumont et al. 2000; Abrassart & Dumont 2001). In the non-LTE regime, the full statistical equation of state for ionization balance is solved during numerical computations of models and all bound-free and bound-bound processes are taken into account along with free-free absorption and scattering. Petrucci et al. (2020) demonstrated that the models of a dissipative warm corona computed by the `TITAN-NOAR` code are able to fully reproduce the temperature stratification inside warm corona and outgoing spectral shape, in agreement with what is generally observed. Here, for the first time, we have tested `TITAN-NOAR` model on a sample of AGN observations to determine the amount of energy dissipated in the warm corona. Once we know how much the warm corona is heated, we are then able to propose a physical mechanism responsible for heating the plasma to the temperatures that correspond to emission in soft X-rays.

There is already a model of soft X-ray excess available for data fitting, named `ReXcor`, which considers emission from dissipative warm corona computed with full transfer of radiation through an ionized skin (Ballantyne 2020), calculated for lamp-post geometry, where intrinsic hard X-rays are produced in the point source above the black hole so that they illuminate a large portion of the innermost disk region (Nayakshin 2000). Then, the X-ray illumination from the outer disk layers is reflected and both radiative heating by photoionization and constant internal heating interplay with each other. Relativistic corrections on the illuminating radiation and outgoing emission are taken into account during final integration over the disk surface. The global energy balance is achieved by assuming a distribution of dissipated energy between the point source, warm corona, and cold disk. In our case, the energy balance is calculated locally only between warm corona and cold disk. Therefore, our model is not tightly bound by the accretion rate and black hole mass and purely estimates the amount of energy dissipated in a warm corona. A complete radiative transfer through the warm corona is computed together with the reflection, but since the warm corona is optically thick, the outgoing emission is devoid of reflection signatures. Thus, our computed model represents only a separate model component responsible for soft X-ray excess ready to use in data fitting packages (e.g., XSPEC).

### 2.1. Energy dissipation in the warm corona

Given the radiative heating is not enough to produce an optically thick, warm layer on the top of an accretion disk (Różańska et al. 2015), we consider that in addition to external illumination, the warm corona is also internally heated; namely, part of the energy generated in accretion process is dissipated in the layer. The example of such heating was considered in the form of heating by magneto rotational instability (MRI), but it was never coupled with spectral calculations (Gronkiewicz & Różańska 2020; Gronkiewicz et al. 2023). Therefore, in the model presented here, we do not specify the origin of this internal heating. However, we do assume that the input energy rate in a warm corona, per unit optical depth and solid angle, $Q$ is uniform, and it corresponds to the flux emitted by the warm corona toward the observer.

$$F_{\rm cor} = 4\pi\tau_{\rm cor}Q, \qquad (1)$$

where $\tau_{\rm cor}$ is the optical depth of the warm corona, which is an input parameter in our numerical calculations. Therefore, total mechanical heating per particle per unit volume is:

$$q_{\rm h} = \frac{4\pi\sigma_{\rm T} Q}{n_{\rm H}}, \quad (2)$$

where $\sigma_{\rm T}$ is a Thomson cross section, and $n_{\rm H}$ is hydrogen density number of the gas. This quantity, after multiplying by electron and proton density number, that is, $n_{\rm H} n_{\rm e} q_{\rm h}$ in erg s$^{-1}$ cm$^{-3}$ is an input parameter into radiative transfer code TITAN.

At each stage of our considerations, we ensure that the energy balance between the warm corona and cold disk is sustained, that is, the total energy deposited via the accretion process is divided between two layers. Taking into account that such energy is converted into radiation (with radiative energy for standard thin disk) and including reprocessing we get:

$$F_{\rm tot} = F_{\rm cor} + F_{\rm int}, \quad (3)$$

where B is the frequency integrated black-body radiation intensity, so $F_{\rm int}$ is the sum of intrinsic emission from disk due to thermalized black-body $4\pi B/(4 + 3\tau_{\rm cor})$ and reprocessed emission $2\pi\tau_{\rm cor} Q$, coming from the base of the corona (Różańska et al. 2015; Petrucci et al. 2020). The above fluxes are needed to estimate the fraction of energy dissipated in the corona relative to the total energy dissipation, $\chi$ (not to be confused with the goodness of fit, $\chi^2_{\rm red}$), which can be derived while fitting numerical models to observations:

$$\chi = \frac{F_{\rm cor}}{F_{\rm tot}} = \frac{\tau_{\rm cor} n_{\rm H} q_{\rm h}}{\sigma_{\rm T} F_{\rm tot}}. \quad (4)$$

Even though we use different symbols here, the internal heating of the warm corona is defined in the same way as in the ReXcor model. Our $\chi$ corresponds to $h_f$ in Eq. 7 from Xiang et al. (2022) and our $q_{\rm H}$ is simply the $\mathcal{H}$ value from that same work. The only difference is that in the ReXcor model, the total energy deposited by accretion is divided into three emitting areas: point X-ray source (lamppost model), warm corona, and cold disk; whereas in our paper, it is divided between the warm corona and cold disk. Therefore, even if our models have different overall geometry connected to an external X-ray source, (i.e., lamppost versus hot inner corona model) the idea of dissipative corona above an accretion disk is the same and we can compare our results directly. In a quantitative sense, we have to keep in mind that the dissipation fraction defined in our model relates to the disk black body radiation intensity and warm corona intensity, expressed as:

$$\chi = \frac{\tau_{\rm cor} Q}{\frac{B}{4+3\tau_{\rm cor}} + \frac{\tau_{\rm cor} Q}{2}}. \quad (5)$$

The total energy flux, $F_{\rm tot}$, generated by the accretion process is directly related to the accretion rate by standard formulae, for which we assume an accretion efficiency parameter. Because the accretion efficiency has long been a topic of discussion among theoreticians and observers alike, for our work here, we do not tie this flux to a particular value of an accretion rate. Our work aims to find observational constraints on the amount of energy released in the warm corona by additional heating. For this purpose, we use radiative transfer code TITAN, which can include different strengths of the illuminating continuum for both sides of the slab and internal heating of the gas. The connection of the observed flux to the eventual accretion rate of the source also comes from observations and we discuss this issue below in this paper.

### 2.2. TITAN and NOAR spectral models

The full description of the procedure of spectral model preparation is given in Petrucci et al. (2020). In this paper, we use this model of the warm, dissipative corona to fit observations of our AGN sample described in Sect. 3. The model grid was computed with the radiative transfer code TITAN (Dumont et al. 2003) coupled with the Monte Carlo code NOAR, where the former accounts for ionization and thermal equilibrium of the gas, and the latter for a detailed treatment of Comptonization. The iteration between both codes undergoes up to convergence and the final angle-dependent spectrum accounts for external X-ray illumination from the top, reflection on the illuminated side, transmission through the gas, illumination by seed photons from the bottom, and additional internal heating constant over gas volume. Throughout the computations, all free-free, bound-free, and bound-bound atomic processes have been taken into account, thereby allowing for the transfer of continuum radiation and lines. The coupling between TITAN and NOAR allows for a complete treatment of the emission from a photoionized, Comptonized medium and can be used in a variety of cases, such as illuminated disk atmospheres (Różańska et al. 2002) and warm absorbers (Różańska et al. 2006) in AGNs. For the present work, the TITAN-NOAR models corresponding to the emission from the warm corona are denoted by a blue unfilled arrow in Fig. 1.

We computed the spectra for a large range of parameters: gas number density, $n_{\rm H}$; warm corona optical depth, $\tau_{\rm cor}$; dissipation rate, $q_{\rm h}$; ionization parameter as the normalization for power-law shaped external X-rays illuminating the surface of the warm corona, $\xi$; power-law spectral index of the illuminated continuum, $\Gamma$; power-law low and high energy cut-offs, $h\nu_{\rm min}$ and $h\nu_{\rm max}$; and the temperature of soft photons injected into the bottom of the warm corona, mimicking the disk black body emission, $kT_{\rm bb}$. Since our goal has been to put constraints on the amount of internal heating of the warm corona, the outcome of radiative transfer calculations is used to build a spectral component that reflects warm corona emission only. We did not fit the reflection and hot corona emission with TITAN-NOAR models. Therefore, for the data fitting described in this paper, we have chosen to keep certain parameter values fixed: $kT_{\rm bb}$ = 7 eV, $\log\xi$=3, $\Gamma$ =1.8, and $n_{\rm H}$=$10^{12}$ cm$^{-3}$, $h\nu_{\rm min}$ = 50 eV, and $h\nu_{\rm max}$ = 100 keV. We only kept the warm corona parameters free when building the table models. These values are typical for AGNs with black hole masses $10^8$ $M_\odot$ (Rees 1984).

The gas density of the warm corona is strongly justified by the model of dissipative, magnetically supported corona above an accretion disk given by Gronkiewicz et al. (2023). The model does not solve energy-dependent radiative transfer and for this reason, it cannot justify the first three parameters that are kept fixed. While the spectral index and ionization parameter have been taken from many previous observational fits (Petrucci et al. 2018), the value of the seed photons temperature is not obvious. Such seed photons are Comptonized in both the hot and warm corona and are responsible for the final spectral shape of the warm corona. With the use of a thermal Comptonization model NTHCOMP (Zdziarski et al. 1996; Życki et al. 1999) available via the X-ray fitting package XSPEC (Arnaud 1996), we show (in Fig. 2) how the final Comptonized spectra depend on the seed photon temperature. Clearly, below the seed photon temperature $kT_{\rm bb}$ = 20 eV, the shape of output spectra is less sensitive to $kT_{\rm bb}$. This limit is equal to the maximum value of $kT_{\rm bb}$ obtained from spectral fitting X-ray/UV data in Petrucci et al. (2018). Hence, it justifies our choice of 7 eV for this parameter in the model.

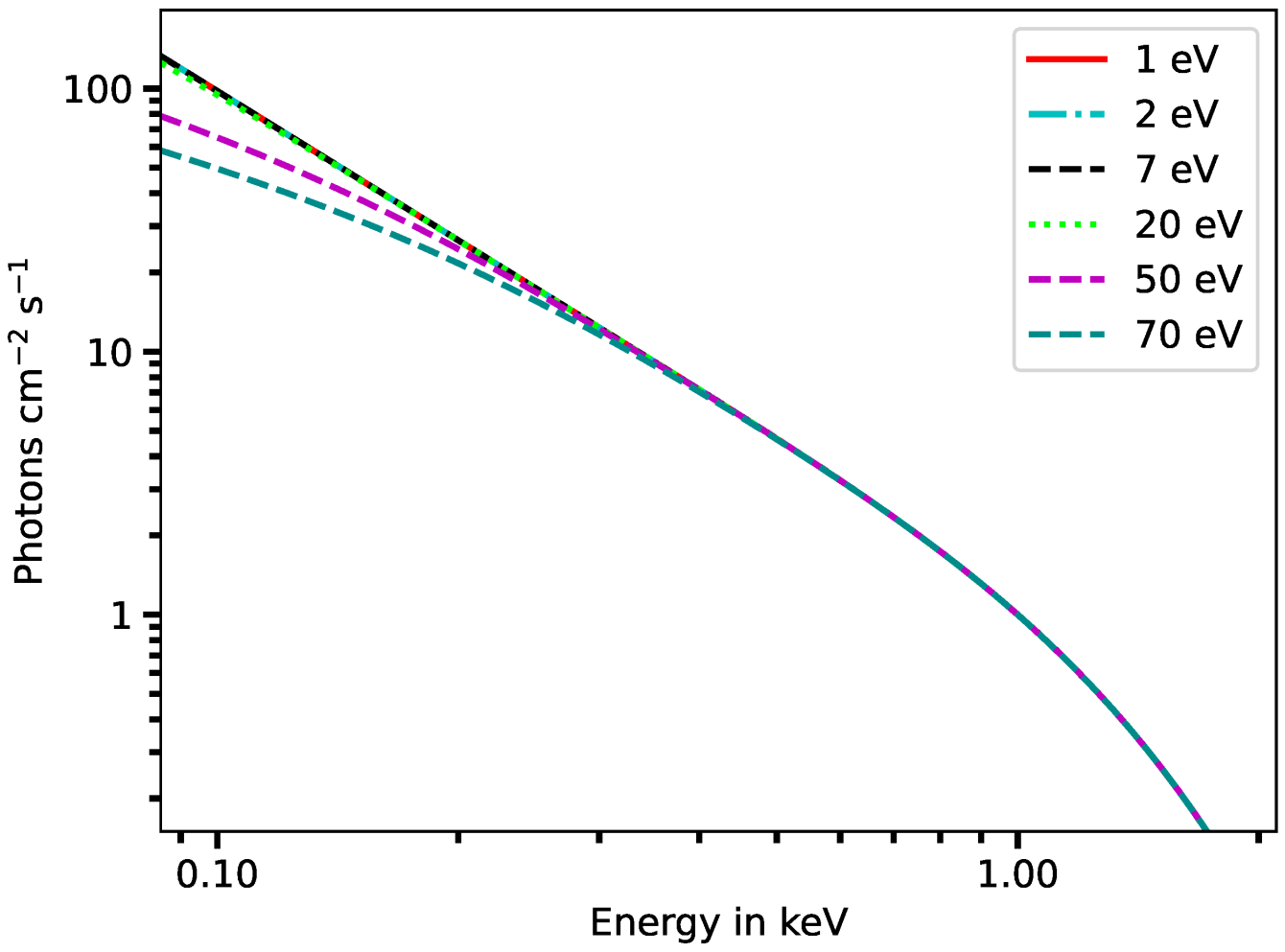

Fig. 2: Dependence of thermal Comptionization computed by NTHCOMP model (see Sect. 2.2 for discussion) on different values of input seed photon temperature ($kT_{\mathrm{bb}}$) ranging from 1 to 70 eV. Below $kT_{\mathrm{bb}}$=20 eV, there is a negligible effect on the shape of the Comptonized spectrum.

With the above assumptions, we are left with only two free parameters, which are the warm corona optical depth, $\tau_{\mathrm{cor}}$, and internal heating of the warm corona, $q_{\mathrm{h}}$. These parameters were varied in the range of [5-30] and [$10^{-23}$-$10^{-21.5}$] erg s$^{-1}$ cm$^3$, while computing our grid of 196 soft X-ray excess models, used below in the data fitting procedure. All TITAN-NOAR models are normalized to the case where the warm corona radius is 10 $R_{\mathrm{g}}$ and the distance to the source is equaled to 10 kpc, marked as $D_{10}$. Then, $R_{\mathrm{g}}$ denotes the gravitational radius defined as $R_{\mathrm{g}}$=$GM_{\mathrm{BH}}/c^2$, where $G$ is the gravitational constant, $M_{\mathrm{BH}}$ is the black hole mass, and $c$ the velocity of light. This means that for a typical black hole mass of $10^8$ $M_{\odot}$, all spectra are diluted by a factor of $(10R_{\mathrm{g}})^2/4\pi D_{10}^2 = 1.82\times10^{-18}$. However, in our analysis, we kept the normalization of TITAN-NOAR model free during the fitting process. In that case, it is interpreted as the surface area of the soft X-ray excess emitting warm corona. Taking the distance to the source $D$ in units of $D_{10}$, they are related as $N_{\mathrm{SE}} = S/D^2$. Thus, assuming a spherically symmetric emitting area, we can estimate the size of the warm emitting region by solving the equation:

$$R_{\mathrm{cor}} = \sqrt{N_{\mathrm{SE}}\,D^2\,(10R_{\mathrm{g}})^2/\pi}, \qquad (6)$$

with appropriate $R_{\mathrm{g}}$ value, calculated using black hole masses from Table 1.

## 3. Sample selection and data reduction

Our AGN sample, drawn from Petrucci et al. (2018, hereafter P18), consisted of 21 nearby AGNs ($z$ = 0.009–0.5725) that have been extensively observed by *XMM-Newton* X-ray telescope. There were a total of 14 type 1 Seyferts, 6 intermediate-type Seyferts, and one quasi-stellar object (QSO). In total, 82 observations were analyzed, which are listed in Table 1, where we also report their activity type, redshift, Galactic absorption, and black hole mass, all taken from previous studies (P18).

The source, PG1114+445 was removed from the original sample due to its strong line of sight obscuration (Serafinelli et al. 2021). On the other hand, the source, NGC 7469, was included in the parent sample. NGC 7469 is a luminous ($L_{\mathrm{Bol}}$=$10^{45}$ erg s$^{-1}$, Behar et al. 2017)), nearby ($z$=0.0162) Sy1 galaxy having strong soft X-ray excess and narrow, neutral FeK$\alpha$ and FeK$\beta$ line components. It has been the subject of several multiwavelength studies that led to the detection of correlated UV/X-ray variability (Kumari et al. 2023) and thermal outflows from the AGN torus (Mehdipour et al. 2018). Furthermore, a soft Comptonization scenario for the production of soft X-ray excess has been repeatedly suggested for this source which makes it an ideal target for testing the dissipative warm corona model.

The sources in our sample were simultaneously observed by *XMM-Newton* EPIC-pn camera (Strüder et al. 2001), along with multiple optical monitor (OM) filters (Mason et al. 2001), with exposure times distributed between ~ 12 and ~ 100 ks. Since the disk black body temperature of the outer disk is fixed in our grid model to 7 eV, the OM data were ignored.

For our analysis, we adopted the Eddington ratios ($L_{\mathrm{Bol}}/L_{\mathrm{Edd}}$) derived in P18, which are reported in the third column of Table D.2 in Appendix D. For the two sources, HB890405-123 and LBQS1228+1116, the black hole mass was unavailable so the accretion rate was undetermined. As shown in Tables 1 and D.2, the sources span a wide range of black hole masses as well as accretion rates. This ensures the possibility of detecting any form of evolution of warm corona properties across masses. The data was reduced by P18 following the standard procedures of Science Analysis Software (Gabriel et al. 2004) guidelines. Also, we decided to omit the spectral data between 1.8–2.4 keV to avoid improper modeling of calibration uncertainties arising from detector quantum efficiency at the Si K-edge (1.84 keV) and mirror effective area at the Au M-edge (~ 2.3 keV) (Cappi et al. 2016). For more details about data reduction, we refer the reader to the P18 paper.

We developed an automatic fitting procedure to fit multiple observations. We use the object-oriented Python interface to the XSPEC called PyXSPEC (Gordon & Arnaud 2021). The spectral files are read confining the energy range to 0.3 -10 keV and ignoring any " bad" channels.

## 4. Warm corona model applied to the AGN sample

All the data have been analyzed using the publicly available XSPEC package (Arnaud 1996), designed for spectral analysis in the X-ray domain. In this section, we formulate the total model used for this paper.

The so-called two-corona approach is most commonly described using dual Comptonization models, treating the hot and warm coronae as disjoint entities radially separated from each other. In the previous work (P18), the authors adopted a phenomenological approach to describe the broadband UV/X-ray data in the context of studying the soft X-ray excess. In the first step, $kT_{\mathrm{bb}}$ was estimated from all the available OM filters with the multicolor disk black body model DISKBB, taking into account the contributions from broad line region (BLR) and galaxy templates. Then the Comptonized emission from warm and hot corona were modeled using the thermal Comptonization model, NTHCOMP (Zdziarski et al. 1996). The $kT_{\mathrm{bb}}$ value estimated from OM part of data was fed into the two NTHCOMP components as input for soft-seed photon temperature to describe soft X-ray excess and hard continuum. The non-relativistic reflection model XILLVER (García & Kallman 2010) was added to fit the Fe K$\alpha$ line and any reflection component present. Finally, all those model components were multiplied by the warm absorber table model (for details see: Cappi et al. 2016) computed with

Table 1: Sample of the 21 AGNs studied in this paper

| Source name | Seyfert activity type | Redshift $z$ | $N_H^{Gal}$ $\times 10^{20}$ [cm$^{-2}$] | log $M_{BH}$ [$M_\odot$] |
|---|---|---|---|---|
| 1H0419-577 | 1.5 | 0.1040 | 1.31 | 8.58 |
| ESO198-G24 | 1.0 | 0.0455 | 3.26 | 8.48 |
| HB890405-123 | 1.2 | 0.5725 | 4.16 | - |
| HE1029-1401 | 1.2 | 0.0858 | 6.28 | 8.73 |
| IRASF12397+3333 | 1.0 | 0.0435 | 1.31 | 6.66 |
| LBQS1228+1116 | QSO | 0.2362 | 2.60 | - |
| MRK 279 | 1.0 | 0.0304 | 1.59 | 7.54 |
| MRK 335 | 1.0 | 0.0257 | 4.09 | 7.15 |
| MRK 509 | 1.5 | 0.0343 | 5.16 | 8.16 |
| MRK 590 | 1.0 | 0.0263 | 2.93 | 7.68 |
| NGC 4593 | 1.0 | 0.009 | 2.05 | 6.28 |
| NGC 7469 | 1.0 | 0.0162 | 5.24 | 7.04 |
| PG0804+761 | 1.0 | 0.1000 | 3.51 | 8.24 |
| PG0844+349 | 1.0 | 0.0640 | 3.22 | 7.97 |
| PG1116+215 | 1.0 | 0.1765 | 1.46 | 8.53 |
| PG1351+640 | 1.5 | 0.0882 | 2.10 | 7.66 |
| PG1402+261 | 1.0 | 0.1640 | 1.44 | 7.94 |
| PG1440+356 | 1.0 | 0.0790 | 1.08 | 7.47 |
| Q0056-363 | 1.0 | 0.1641 | 1.94 | 8.95 |
| RE1034+396 | 1.0 | 0.0424 | 1.36 | 6.41 |
| UGC 3973 | 1.2 | 0.0221 | 6.61 | 7.72 |

**Notes:**
Redshifts are taken from NASA Extra-galactic Database- NED.
Galactic column densities $N_H^{Gal}$ are taken from Willingale et al. (2013).
The black hole masses are adopted from Petrucci et al. (2018) and see Bianchi et al. (2009) for detailed references.

Table 2: Free and fixed parameters used in the total composite model.

| Model Component | Parameters: Free | Parameters: Fixed |
|---|---|---|
| TBABS | – | $N_H^{Gal}$ |
| CLOUDY | $\xi_{CL}$, $N_H^{CL}$ | $v_{turb}$ = 100 km s$^{-1}$ |
| TITAN-NOAR | $q_h$, $\tau_{cor}$, $N_{SE}$ | $kT_{bb}$ = 7 eV, log $\xi$=3, $\Gamma$ =1.8 |
| NTHCOMP | $\Gamma_{hc}$, $N_{hc}$ | $kT_{hc}$ = 100 keV, $kT_{bb}^{hc} = kT_{bb}$ |
| XILLVER | $N_X$ | $\Gamma_X = \Gamma_{hc}$, $E_c$ = 300 keV, $\xi_X$ = 1 |
| Sum | 8 | 10 (2 pairs are tied) |

the use of publicly available CLOUDY code (Chatzikos et al. 2023, and references therein).

We took a similar approach, namely, by replacing one of the thermal Comptonization components responsible for soft X-ray excess emission, with the TITAN-NOAR warm corona emission model described in Sect. 2. The resulting total composite model is then:

$$\text{TBABS} \times \text{CLOUDY} \times (\text{TITAN-NOAR} + \text{NTHCOMP} + \text{XILLVER}), \quad (7)$$

where the thermal Comptonization model, NTHCOMP accounts for the hot corona emission, and the warm corona emission is described by TITAN-NOAR model. Apart from the normalization $N_{SE}$, the TITAN-NOAR model has two free parameters: warm corona optical depth $\tau_{cor}$, and total internal heating $q_h$. The TITAN code computes temperature structure in the optically thick warm corona, so we do not get any singular estimate of warm corona temperature from our model. As described in Sect. 2, TITAN provides the value of internal heating of a warm layer existing on top of a cold disk and also computes the optical depth of a warm corona, which is not given by NTHCOMP. Since the TITAN models were generated by assuming a $kT_{bb}$= 7 eV, the input for seed photon temperature for hot corona component ($kT_{bb}^{hc}$) was kept frozen at 7 eV for the entire fitting process. For this reason, we excluded OM data from our analysis. The remaining parameters were kept the same as in P18.

We fixed the temperature of hot corona $kT_{hc}$ = 100 keV, high energy cutoff ($E_c$) of XILLVER illuminating continuum to 300 keV. Both of these constraints have negligible effect on the inferred parameters when fitting data below 10 keV. Assuming reflection from distant cold matter, we also fixed the ionization parameter (log $\xi_X$) of XILLVER to zero. The spectral photon index of the hot corona and the reflection, $\Gamma_{hc} = \Gamma_X$ were tied to achieve a more consistent picture of the modeling. Finally, the solar iron abundance ($A_{fe}$ = 1) and disk inclination ($i$=30°) were assumed for the reflection model. Again, these choices have a negligible effect on the key expected result of our analysis, which is an interplay between two model components, TITAN-NOAR and NTHCOMP responsible for soft X-ray excess and hot corona emission, respectively.

The neutral Galactic absorption along a line-of-sight was accounted by TBABS model, where the $N_H^{Gal}$ value (consisting of a contribution from both ionized and molecular hydrogen) was frozen at the values taken from Willingale et al. (2013), which are given in fourth column of Table 1. Following P18, the remaining residuals in the soft band were modeled with a warm absorber component given by CLOUDY table model (same as in P18). It has two free parameters: the ionization parameter, $\xi_{CL}$, and column density, $N_H^{CL}$. The turbulent velocity, $v_{turb}$, was fixed at 100 km s$^{-1}$. In summary, there are eight free parameters in the total model, including the normalization of additive model components, $N_{SE}$, $N_{hc}$ for the hot corona, and $N_X$ for reflection. All of these are listed in Table 2.

For completeness, we tested the effect of replacing the XILLVER model with the relativistic model RELXILL (García et al. 2013, 2014) on a prototypical source: 1H0419-577. The detailed impact of the warm corona is discussed in Appendix C. We found that the relativistic reflection only contributes slightly to the soft X-ray excess emission and has negligible impact on the warm corona properties. Moreover, some physical parameters have not been constrained, such as the radius of the inner accretion disk, iron abundance, and the black hole spin. Therefore, the non-relativistic reflection XILLVER model was ultimately used in our analysis, allowing us to fit the reflection features reasonably while, at same time, keeping the total model on the lowest level of complication.

## 5. Results

### 5.1. Goodness of fits

The final reduced chi-squared fit statistic $\chi^2_{red}$ ($\chi^2_{stat}$/d.o.f. – per degrees of freedom) for each data set is presented in the second column of Table D.2. In our analysis, 98% of the observations have $\chi^2_{red}$ < 1.5. The few datasets corresponding to larger $\chi^2_{red}$ values have been re-analyzed individually. In most cases, these were the ones with complex absorption below 2 keV. For instance, MRK 509, ObsID: 0601390801 required the addition of an extra warm absorber, as well as the freeing of the ionization parameter $\xi_X$ of XILLVER model. It improved this particular data fit from 354.79/219 to 261.33/216 ($\Delta\chi^2_{\rm stat} \sim 93$ for three extra d.o.f). A relative likelihood value of $\sim 10^{-37}$ based on the Akaike information criterion (AIC) indicates that the second model is much more likely to correctly describe the data.

With the exception of a few sources with a high value of column density, adding an extra warm absorber has a negligible effect on the inferred optical depth, internal heating of warm corona, and hard X-ray photon index. Four sources, Mrk 509, RASF12397+3333, UGC 3973, and RE1034+396, required two warm absorbers to completely describe the complex features below 2 keV. Details of the distribution of warm absorber column density and ionization level in our sample are discussed in Appendix A. Finally, in the case of Mrk 335, the fit was greatly improved after adding a partial covering absorption model, by a neutral medium ZPCFABS (available in XSPEC), with $\sim 90$ % covering (see Longinotti et al. 2013), after leaving $\xi_X$ free during the fitting procedure (see Appendix D, Table D.1 for best-fit values). The significant improvement in statistics is supported by a large change in $\chi^2_{\rm stat} \sim 250$ for three extra d.o.f and, hence, a decrease in the AIC value by 239.

To test the case of mild ionized reflection, we refitted all observations, keeping the ionization parameter of XILLVER unfrozen during the fit. The majority of sources turned out to have a neutral Fe K line and, in some cases, the value of $\xi_X$ could not be constrained. Unfreezing $\xi_X$ did not have a significant effect on other parameters, except that their parameter uncertainties were larger. Hence, we decided to keep it frozen to 1 for further discussion of the final results.

Furthermore, we had a closer look at the cases with reduced Chi-square values between 1.3 and 1.5, checking the goodness of fit manually. First, we identified a set of ten observations corresponding to five AGNs (1H0419–577, MRK 335, NGC 4593, NGC 7469, and RE1034+396), which exhibited narrow residuals at $\sim 0.40 - 0.90$ keV. These residuals were identified as narrow emission lines from their corresponding high-resolution reflection grating spectrometer (RGS) data, originating in the narrow line region (NLR) (further details on the identification of emission lines in these sources can be found in Steenbrugge et al. 2003; Di Gesu et al. 2013; Longinotti et al. 2013; Behar et al. 2017). Usually, they correspond to a mixture of nitrogen, oxygen, and iron lines (Bianchi et al. 2006; Middleton et al. 2009; Ebrero et al. 2010) that can be studied in detail with high-resolution instruments ($\sim 5$ eV) such as micro-calorimeters (e.g., X-IFU) available with future missions, including ATHENA (Barret et al. 2023). Secondly, we identified another distinct set of 11 observations corresponding to four AGNs (1H0419-577, MRK 335, NGC 4593, and NGC 7469) for which, in addition to the above-mentioned residuals, the warm corona internal heating was extremely low (i.e., close to the boundary of parameter space). Checking the overall distribution of statistics for the second set, we found that all of them displayed two local minima, with one of them always located close to the edge of TITAN-NOAR parameter space.

In our modeling, we refit the above-mentioned residuals using narrow Gaussians with the width $\sigma$ frozen at a limit of 1 eV and energy centroid at roughly the peak value of sharp residual feature. As an example, we show in Fig. 3 the spectral decomposition of MRK 279 data (ObsID: 0302480401), without (upper panel) and with one Gaussian (bottom panel) added. This allows us to improve fit statistic by 0.17 dex without a drastic change of warm corona parameters. The significance of the improvement in statistics was supported by a reduction of 37 units in AIC value. It must be noted that refitting the second set of 11 observations (mentioned above) with additional narrow Gaussian lines improved all the fits in two ways: by lowering the reduced chi statistic and removing the local minimum located close to the limit of the parameter range. A better fit was found with a noticeable change in both the warm corona parameters and only a minor change in other parameters. An example of such improvement is presented in Fig. 4, where the upper panel shows spectral decomposition of 1H0419-577 data (ObsID: 0148000201) fitted with our basic model given by Eq. 7, and lower panel shows the same model with two narrow Gaussians at energies 0.59 keV and 0.90 keV added, which caused the increase in heating in the warm corona, $\log q_{\rm h}$, by 0.66 dex and $\tau_{\rm cor}$ from 5 to 16.7 (Tab. D.2). From the RGS observations, those lines were interpreted as O VII (0.56 keV) and Ne VII (0.90 keV) emission lines from NLR (Di Gesu et al. 2013). All the fit improvements shown above are taken into account in Table D.2 and in the results presented below.

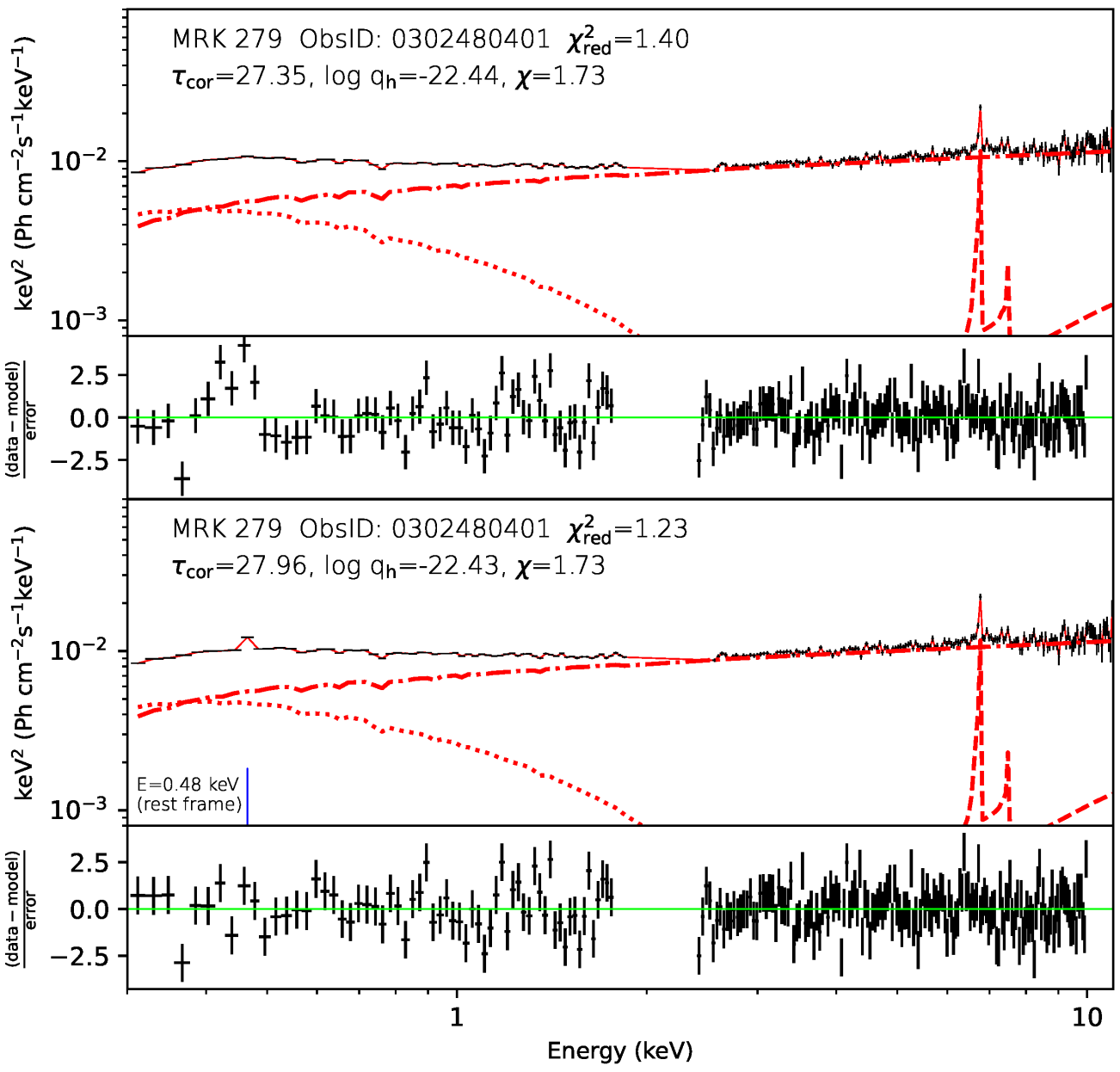

Fig. 3: Spectral model decomposition of MRK 279 data set with ObsID: 0302480401 fitted with our basic model given by Eq. 7 – upper panel and the same model with one narrow Gaussian added - lower panel. Adding one narrow line allowed us to improve the fit statistics without affecting the warm corona marked by a dotted, red line. NTHCOMP component is presented by the dotted-dashed red line, XILLVER is shown by the dashed red line, while observations are given by black crosses. For each case, the ratio of data to the folded model is plotted below.

### 5.2. *Properties of the warm corona*

Here, we present the results of our data fitting procedure, namely: the best-fit model parameters and correlations between them, followed by an interpretation of those results. The values of each fitted parameter are presented in Table D.2. A list of remaining fit parameters is deferred to Table D.3 in Appendix D. After fitting a self-consistent model of a warm, dissipative corona, cooled by Compton scattering, we constrained the internal heating responsible for the active, warm layer together with its optical depth.

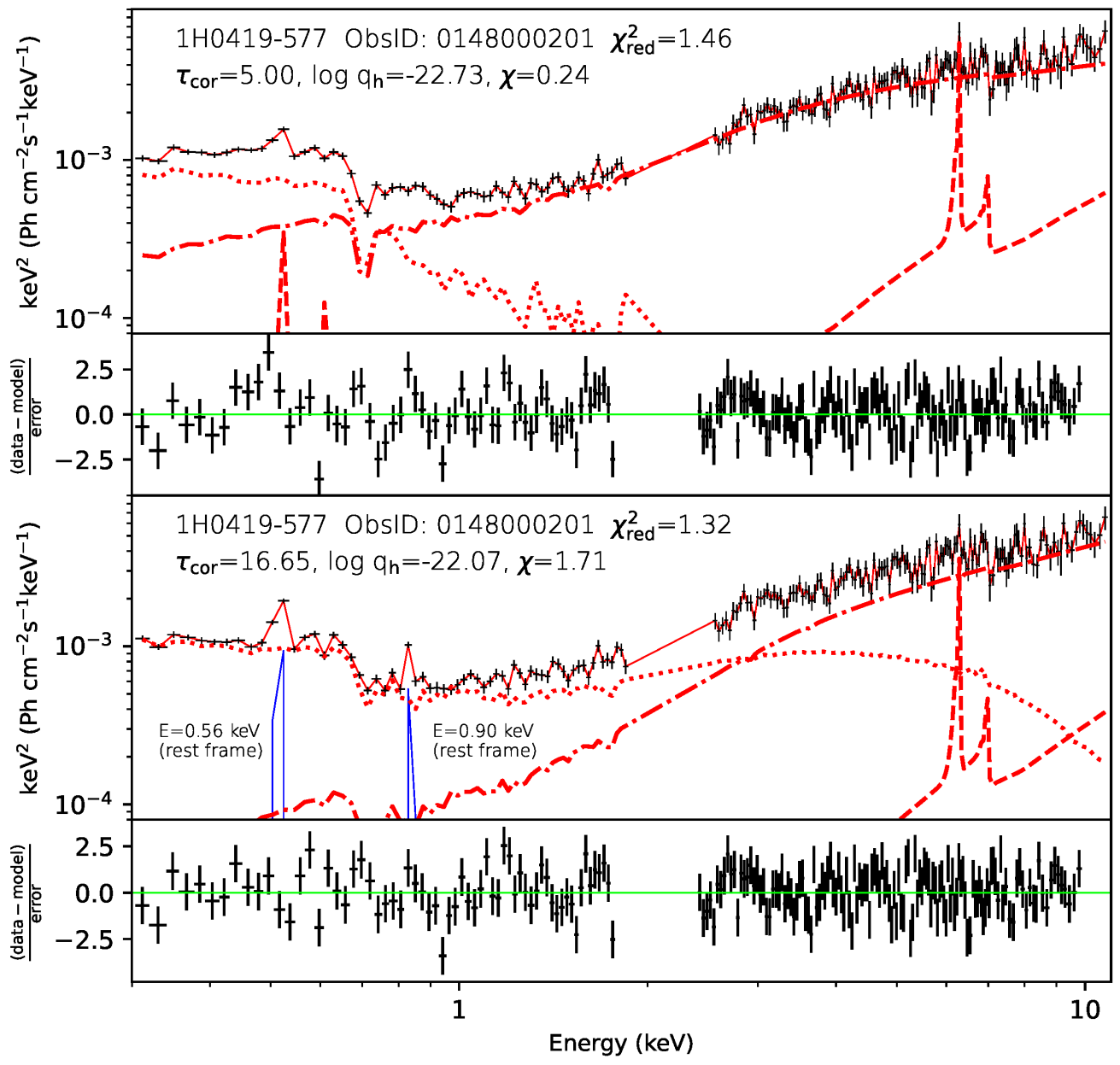

Fig. 4: Spectral model decomposition of 1H0419-577 data set with ObsID: 0148000201 fitted with our basic model given by Eq. 7 (upper panel) and the same model with two narrow Gaussians added (lower panel). Adding two narrow lines with fixed widths allowed us to obtain a global minimum with a relatively strong warm corona marked by a dotted, red line. NTHCOMP component is presented by a dotted-dashed red line, XILLVER is shown by a dashed red line, while observations are given by black crosses. For each case, the ratio of data to the folded model is plotted below.

Subsequently, adopting Eq. 5, we computed the dissipation fraction of the warm corona, $\chi$ (the values are listed in the eighth column of Table D.2).

In the fourth column of Table D.2, we present the hardness ratio (HR), considering count rates and adopting the prescription: H (hard photon counts in range 2.0 - 10 keV) and S (soft- photon counts 0.3 - 2 keV). The hardness ratio is calculated as:

$$\mathrm{HR} = \frac{\mathrm{H} - \mathrm{S}}{\mathrm{H} + \mathrm{S}}. \tag{8}$$

We are aware that HR given as a ratio of measured counts depends on the effective area of an instrument. The *XMM-Newton* EPIC-pn has a higher effective area in the soft X-ray range compared to spectra above 2 keV, which means that for a similar flux level, at both hard and soft energy ranges, the spectra will always be softer. Extreme values of HR may only be interpreted as truly hard or truly soft sources. Values in the middle are better described by spectral fitting. Due to constraints put on the hard power law photon index, this value is not directly indicative of the strength of the hot corona.

While the HR values computed in this work do not directly demonstrate any connection to the canonical states in X-ray binaries (XRBs; Motta et al. (2009); Dunn et al. (2010)), it does indicate at least qualitatively, whether emission from the hot corona dominates over the warm corona and vice versa. Hence, it is a model-independent way to capture the 'state' of an inner accretion disk.

The total unabsorbed flux ($F_{0.3-10}$) in the energy range of 0.3–10 keV was obtained by convolving the CFLUX XSPEC model with the group of additive components in our composite model given by Eq. 7. These fluxes were used to estimate the strength of soft X-ray excess (SE) as:

$$\mathrm{SE} = \frac{F_{0.3-2}(\texttt{CFLUX*TITAN-NOAR})}{F_{0.3-2}(\texttt{CFLUX*NTHCOMP})}. \tag{9}$$

It must be noted that multiplying a CLFUX does not affect the best-fitted parameter values of the model. It merely scales the flux array, calculated from all the components it acts upon between the given energy points and returns the integrated flux with its error. The strength of soft X-ray excess and total unabsorbed flux are listed in ninth and tenth column of Table D.2, respectively.

We divided our sample into three groups based on the fitted values of the hot corona photon index. Hard sources having $\Gamma_{hc} < 1.70$, intermediate for $1.70 < \Gamma_{hc} < 2.00$, and soft for sources with $\Gamma_{hc} > 2.0$. The total number of observations in these three categories are 32, 34, and 16, respectively. In Fig. 5, we also present the statistical distribution of those three groups over warm corona-fitted parameters. A significant fraction of observations in our sample agreed to moderately large optical depths of warm corona (Fig. 5, left), with a mean of $\tau_{cor}$= 18.26 ± 0.12. In Fig. 5, (right) the internal heating of the warm corona $\log q_h$ is also presented, and it shows two peaks around $\log q_h$ ∼ -22.50 and ∼ 22.00. The mean value of $\log q_h$= -22.25 ± 0.03. Comparing it with the distribution of warm corona optical depth (on the left), we noticed that hard sources (blue dash-dotted line) are associated with warm corona having lower optical depth, but stronger internal heating. The inverse is seen in soft sources (red solid line). Intermediate sources (green dashed line) have a wide spread in $\tau_{cor}$ values but at slightly lower internal heating values.

Such behavior of warm corona properties with $\Gamma_{hc}$ prompted us to check for correlations between warm corona parameters and spectral index of hard X-ray radiation as presented in Fig. 6 for all 82 observations. Each of the 21 sources is represented by markers of different styles, referenced in the bottom right of Fig. 6. This convention is maintained throughout the paper. Markov chain Monte Carlo (MCMC) simulations showing the distribution of Pearson's r-values and p-values are presented in Appendix B.

As seen from Fig. 6 (b), based on all observations, we obtained a mild positive correlation between $\tau_{cor}$ and hot corona spectral index $\Gamma_{hc}$, with a Pearson's rank correlation (r-value) **of** 0.38 ± 0.06, which corresponded to ≫ 99 % confidence level (see Appendix B for details). The grey dashed line in this figure indicates the best fit linear regression obtained using the bivariate correlated errors and intrinsic scatter (BCES) technique (Akritas & Bershady 1996; Nemmen et al. 2012). We considered the orthogonal least squares condition, which treats both variables independently and also takes into account the effect of variable uncertainties. The equation of best fit line is represented as $\tau_{cor} = (48.70 \pm 11.63)\,\Gamma_{hc} - (68.15 \pm 20.39)$, indicating a very steep dependence of $\tau_{cor}$ on $\Gamma_{hc}$. Although it can be seen that individual sources do not necessarily follow a positive trend, the overall tendency of our sample dictates that steepening of $\Gamma_{hc}$ is associated with higher $\tau_{cor}$. The observed correlation between $\tau_{cor}$ and $\Gamma_{hc}$ could be explained by the fact that harder X-rays from a hot corona illuminating the warm corona above a cold disk can give rise to radiative or thermally driven outflow from the outer warm corona layer, which reduces its optical depth. The explanation proposed above has to be justified by multiwavelength variability studies.

Furthermore, we explored the distribution of r-value between $\log q_h$ and $\Gamma_{hc}$. Considering the scatter of points, we obtained a Pearson's r-value centered at -0.37 ± 0.08 corresponding to a confidence level greater than 99% (see Appendix B for details). The slope of best-fit line (shown as grey dashed line Fig. 6 (a)) is again quite high, $\log q_h = (-1.20\pm0.40)\,\Gamma_{hc} - (20.09\pm0.73)$. The large dispersion observed in this figure mainly corresponds to one observation of ESO198-G24, PG0844+349, and UGC 3973. They are also explicitly annotated in Fig. 6. The particular observation of ESO 198-G24 (ObsID: 305370101) contributing to dispersion, has the largest soft X-ray excess strength ($\sim 14$, see Table D.2) of the entire sample. The source PG0844+349 is a case of a strong line of sight absorption ($\gtrsim 10^{22}$ cm$^2$), which is not adequately modeled using our simplistic warm absorber model. The source UGC 3973 is a prototypical source with an almost negligible reflection component (Gallo et al. 2011). This can also be observed from our analysis, where this source's soft X-ray excess strength is one of the highest and thereby dominating the entire spectrum. The one particular observation (ObsID:400070301) contributing to deviation has a large soft excess strength and an extremely high dissipation factor, which could cause dispersion in our scatter diagram. Nevertheless, a higher warm corona heating corresponds to the harder hot corona emission, which may suggest a common mechanism responsible for heating both coronae.

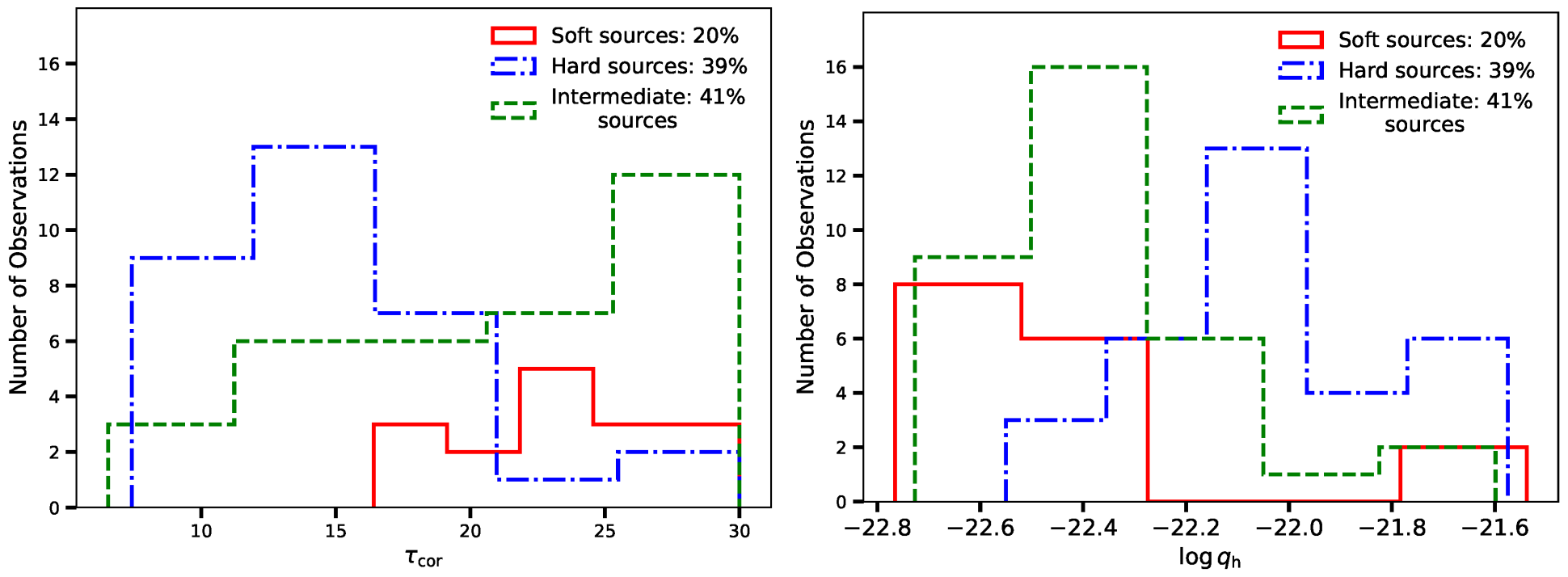

**Fig. 5.** Distribution of the warm corona optical depth, $\tau_{cor}$ (left), and internal heating, $\log q_h$ (right), from fitting the total model described in Sect. 4. Blue histogram denotes sources with $\Gamma_{hc} \leq 1.7$, red – $\Gamma_{hc} \geq 2$ (soft) and the rest are denoted by green. The percentages denote the fraction of the total sample consisting of soft, hard, or intermediate-type sources.

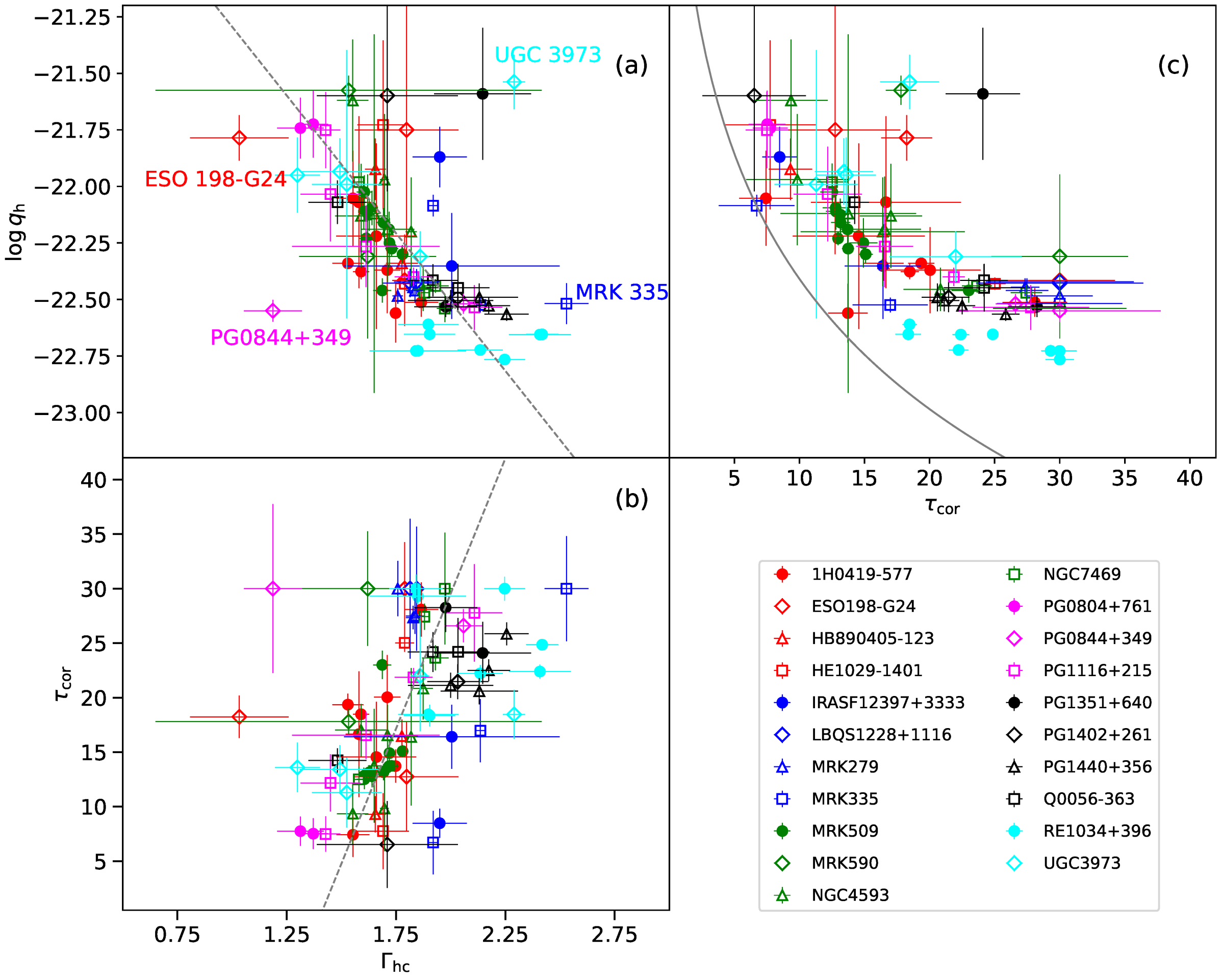

Fig. 6: Scatter plots showing the relationship between internal heating ($\log q_h$) inside warm corona with hot corona spectral index $\Gamma_{hc}$ (a), optical depth ($\tau_{cor}$) of warm corona with $\Gamma_{hc}$ (b), and variation of $\log q_h$ vs. $\tau_{cor}$ (c). The best fit linear regression is indicated by a grey dashed line. The grey solid curve in the top-right panel corresponds to $\chi = 1$ (Eq. 5), marking the boundary between active $\chi < 1$, and passive disk with $\chi > 1$. Different symbols associated with sources are given in the bottom right.

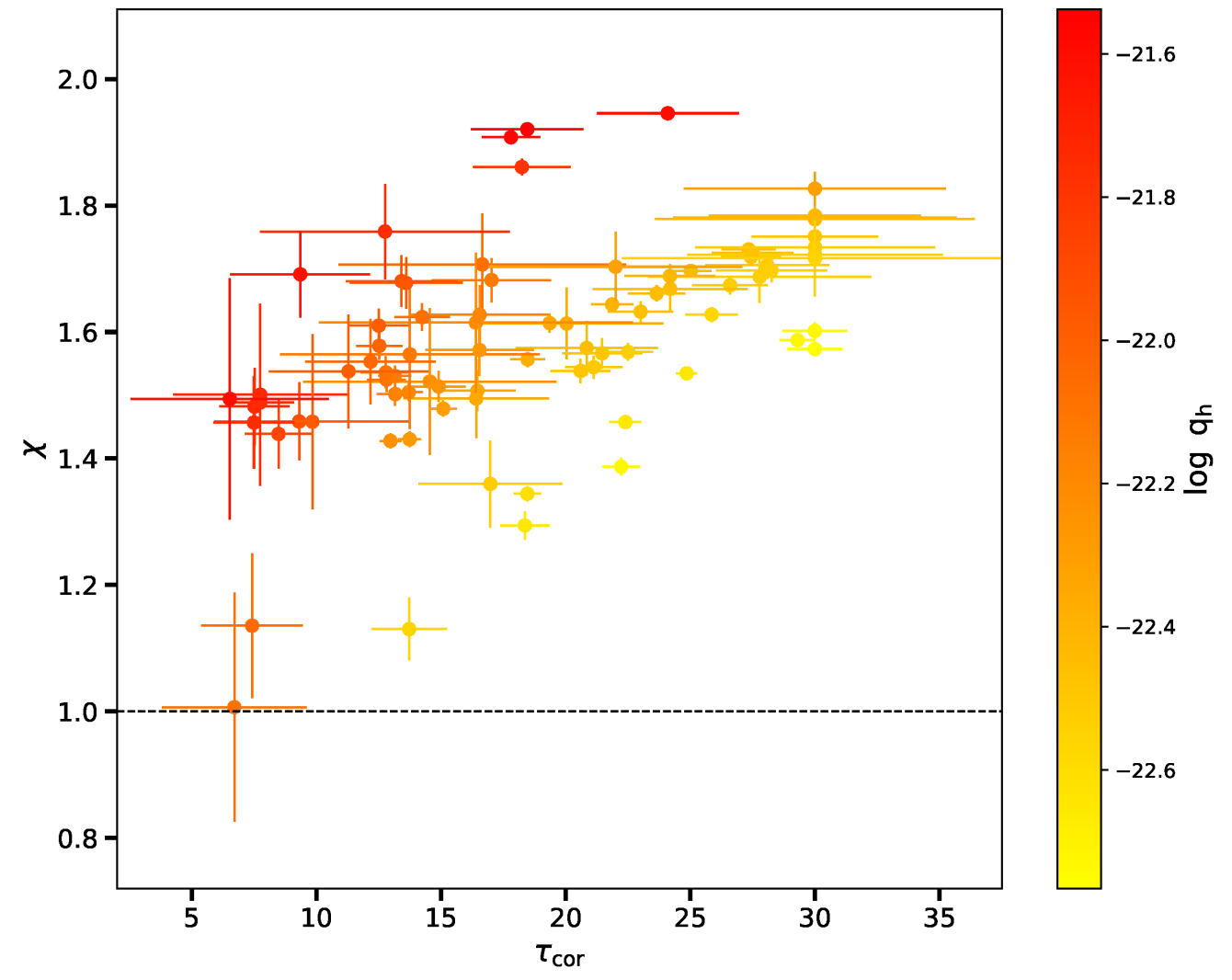

Fig. 7: Scatter plot showing the dissipation factor $\chi$ against warm corona optical depth $\tau_{\rm cor}$ color-coded by the total internal heating value shown on the side color bar. Black dotted line at $\chi$=1.0 denotes the transition region between passive disk – $\chi$>1, and active disk – $\chi$<1.

Judging by the variation between optical depth with total internal heating ( Fig. 6 (c)), there is an inverse non-linear relation between these two parameters. The grey solid line in the top right panel represents the boundary between the passive disk with $\chi > 1$ and the active disk with $\chi < 1$, along with two observations (corresponding to 1H0419-577 and Mrk 335), which are located very close the boundary. The source PG1402+216 is also close to the boundary upon considering error bars. These may indicate the possibility of a transition between the above two configurations. While most observations are well within the dissipative warm corona regime on top of a passive disk, we notice as the optical depth decreases, the data points approach the $\chi$=1 line.

The dependence of dissipation rate $\chi$ on warm corona optical depth and total internal heating is shown in Fig. 7. Extremely high values of $\chi$ were observed in a few cases for sources having large $\tau_{\rm cor}$ and $\log q_{\rm h}$. On the other hand, some points lie below $\chi$= 1.2. These usually have $\tau_{\rm cor}$ < 15 and $\log q_{\rm h}$ > -22.4. Comparing with Fig. 5 (right) and Fig. 6 (b), these belong to a small subsample of observations falling within the intermediate range of $\Gamma_{\rm hc}$ values of 1.7 – 2.0, which suggest a less active warm corona. Hence, the standard accretion disk is actively producing seed photons which can lead to efficient cooling of hot corona as well, leading to overall softening of the photon index. Nonetheless, the majority of points are confined to $\chi$ values between 1.2–1.7, which follows the computations performed by Petrucci et al. (2020, see Fig.2) and earlier estimates of warm corona properties (P18), requiring a large amount of energy dissipation inside warm corona to reproduce the soft X-ray excess.

Next, we compared the fraction of energy dissipated inside the warm corona against the HR, calculated using Eq. 8. As shown in Fig. 8 (left), the observations populate the top two quadrants almost equally, with few points close to the $\chi$=1 line. By looking at the points restricted to the given source, we see that the amount of energy dissipated in warm corona depends on the epoch of observation without a large change in HR. It suggests that even though the rate of dissipation changes inside the warm corona, the relative emission from the hot and warm corona is unaffected. It may be possible that a common source of mechanical heating for both coronae can influence them in such a way as to strike a balance between their emission and maintain similar HRs.

A similar distribution of $\chi$ versus HR, now color-coded by the total internal heating of warm corona is shown in Fig. 8 (right), where we see that the points corresponding to high values of heating seem to be restricted close to HR $\approx$ 0. Those cases are also associated with lower values of warm corona optical depth and low (hard) values of $\Gamma_{\rm hc}$ as evident in Figs. 6 and 7, respectively. While the HR values are not indicative of $\Gamma_{\rm hc}$, it does suggest that comparable emissions from both hot and warm corona could be heavily influenced by stronger heating.

Such a dominance exhibited by the hot corona could indicate the onset of some strong outburst from the inner region of the disk. As shown by Körding et al. (2006), quasars with high radio loudness occupy the hard state and hard-intermediate state in the disk fraction luminosity diagrams (DFLDs). However, a direct comparison of Fig. 8 with DFLD is not possible since we are not taking into account the energetics of an accretion disk; however, the coincidence of high heating values associated with 'HR $\approx$ 0' is a quite striking feature. Due to HR being slightly biased towards soft X-ray energies (as discussed at the beginning of Sect. 5.2), it is possible that observations close to a zero value of HR correspond to harder states (HR > 0). As seen from the figure, those states correspond to high internal heat dissipation inside the warm corona. Strong outflows launched from the disk can also be the cause of the destruction of the warm corona, explaining the smaller values of optical depth.

### *5.3. Warm corona in different types of AGNs*

For our further analysis, we adopted the black hole masses and accretion rates from P18 and Bianchi et al. (2009, and references therein). For almost all sources, the black hole mass has been estimated and is given in the last column of Table 1. Also, the Eddington luminosity to bolometric luminosity ratio that directly corresponds to the accretion rate has been evaluated and we list this quantity in the third column of Table D.2.

Fig. 9 displays warm corona properties as a function of accretion rate and black hole mass, the latter represented by a color bar. The overall conclusion is that properties of the warm corona do not depend on the accretion rate, where $\chi$ and $\tau_{\rm cor}$ remain on the same level across a wide range of Eddington ratios. We observe almost vertical dependence of $\chi$ and $\tau_{\rm cor}$ for the sources at the same black hole mass. Only minute changes may be noticed when looking at the source of a given accretion rate. This may be caused by thermal instability that was recently predicted by Gronkiewicz et al. (2023). The dissipation fraction, $\chi$, displays a mild variation, decreasing from ~ 1.95 to 1.40 with an increasing accretion rate. Due to the coupled behavior of the standard disk with the warm and the hot coronae, to fully capture the evolution of warm corona in AGNs, it is necessary to have multiwavelength data across epochs where the AGN might have undergone large changes in accretion rate or total flux.

A noticeable decrease in the hardness ratio with increasing accretion rate is observed (right upper panel of Fig. 9), indicating that emission from the warm corona becomes stronger relative to the hot corona emission. This is also observed in the increase in SE with accretion rate (bottom-right plot of Fig. 9). Hence, for highly accreting sources, the low value of the hardness ratio is accompanied by the higher value of soft X-ray excess strength. Studying these trends with a larger sample available in the future may indicate the importance of the warm corona in bright sources

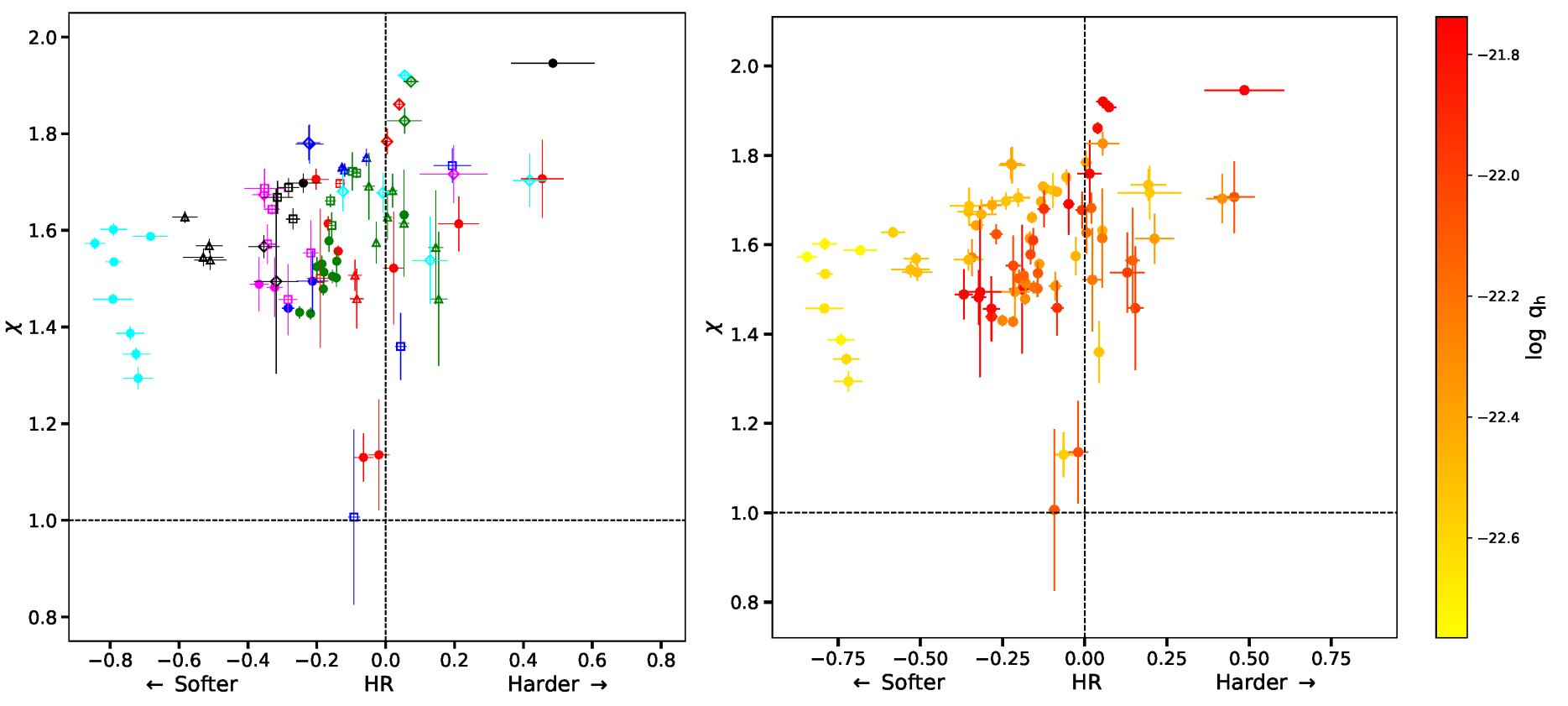

**Fig. 8.** Plots showing hardness ratio versus dissipation factor divided into four quadrants with dashed lines given by HR=0 and $\chi$ = 1, the latter being a transition between passive and active disk. Different symbols associated with sources are the same as in Fig. 6.

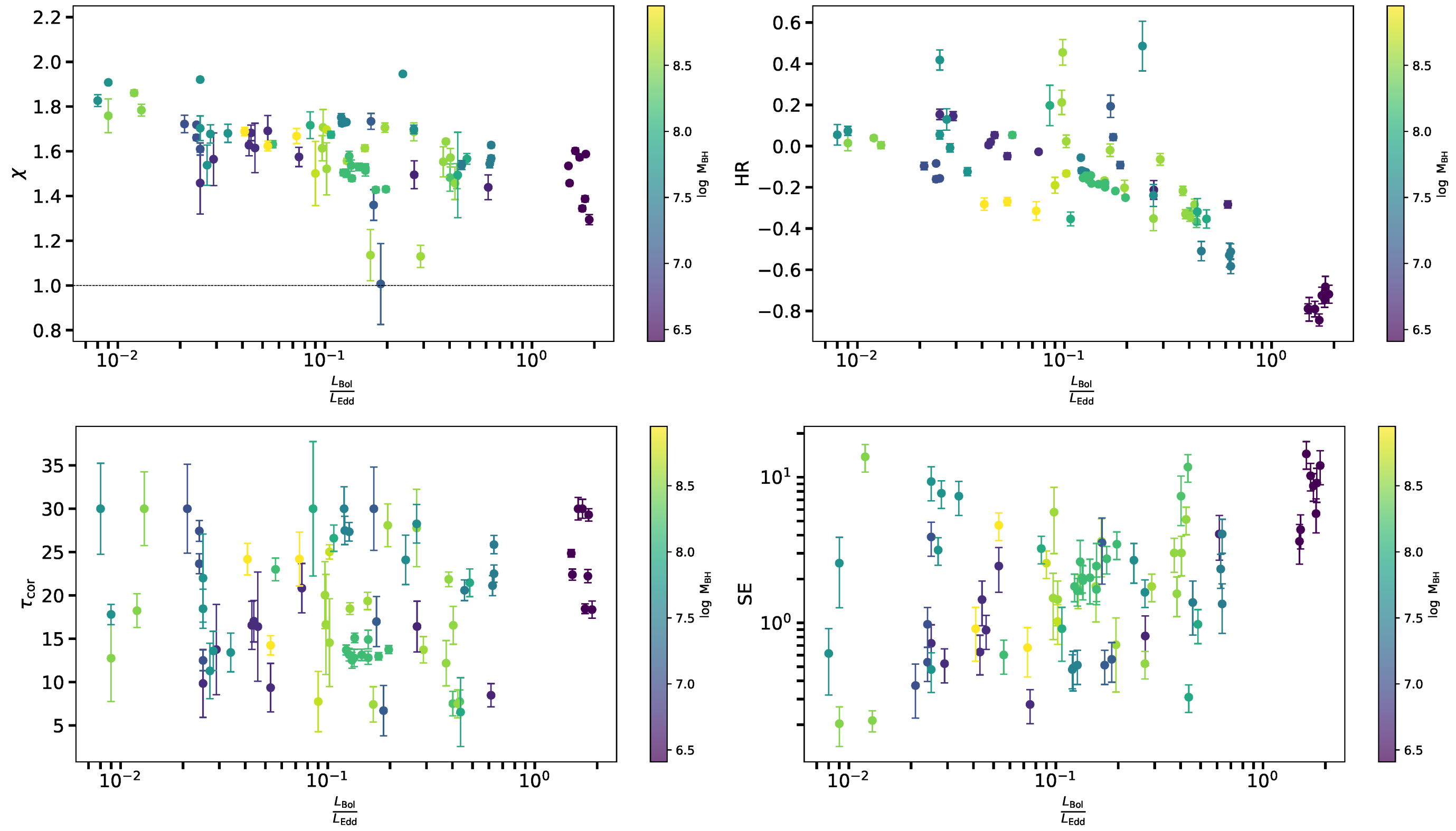

Fig. 9: Dependence of warm corona parameters on accretion rate taken from P18 (left panel) and soft excess strength and hardness ratio (right panel). The color map represents the black hole mass of various AGNs. Two sources-HB890405-123 and LBQS1228+1116 are not included in these figures due to the unavailability of measurement of black hole mass.

and the connection of warm corona dissipation rate with the accretion process.

### *5.4. Extent of the warm corona*

As derived in Sect. 2.2, the normalization of `TITAN-NOAR` model, $N_{\rm SE}$, can be related to the size of the warm corona, according to Eq. 6. Here, the extent of a warm corona refers to the outer radius of a warm corona. We estimated the radii for 19 sources with available distance measurements, totaling 78 observations. The resulting radius from warm corona normalization, together with the distances to the source, is shown in Table 3. Only a few sources, namely, IRASF12397+3333, MRK 335, MRK 509, MRK 590, NGC 4593, NGC 7469, PG0844+349, and UGC 3973, have had redshift-independent distances measured by an AGN dust reverberation mapping technique. For the remaining sources, we used the Hubble distance taken from NED (NASA/IPAC Extragalactic Database), with a Hubble constant $H_0$=68.7 km s$^{-1}$ Mpc$^{-1}$.

The dependence of warm corona parameters, such as the optical depth, internal heating, dissipation fraction, and soft excess strength, together with observed hardness ratio, on the warm corona radius is shown in Fig. 10. The extent of warm corona ranges between ~ 7– 408 $R_{\rm g}$, in agreement with recent works (Kubota & Done 2018; Zoghbi & Miller 2023; Porquet et al. 2024; Vaia et al. 2024). The large values may be slightly overestimated due to the black hole spin being neglected (Porquet et al. 2024); however, $R_{\rm cor}$ values up to $10^2 R_{\rm g}$ were recently reported by (Porquet et al. 2024) based on the `ReXcor` and `RELAGN` (Hagen & Done 2023) models.

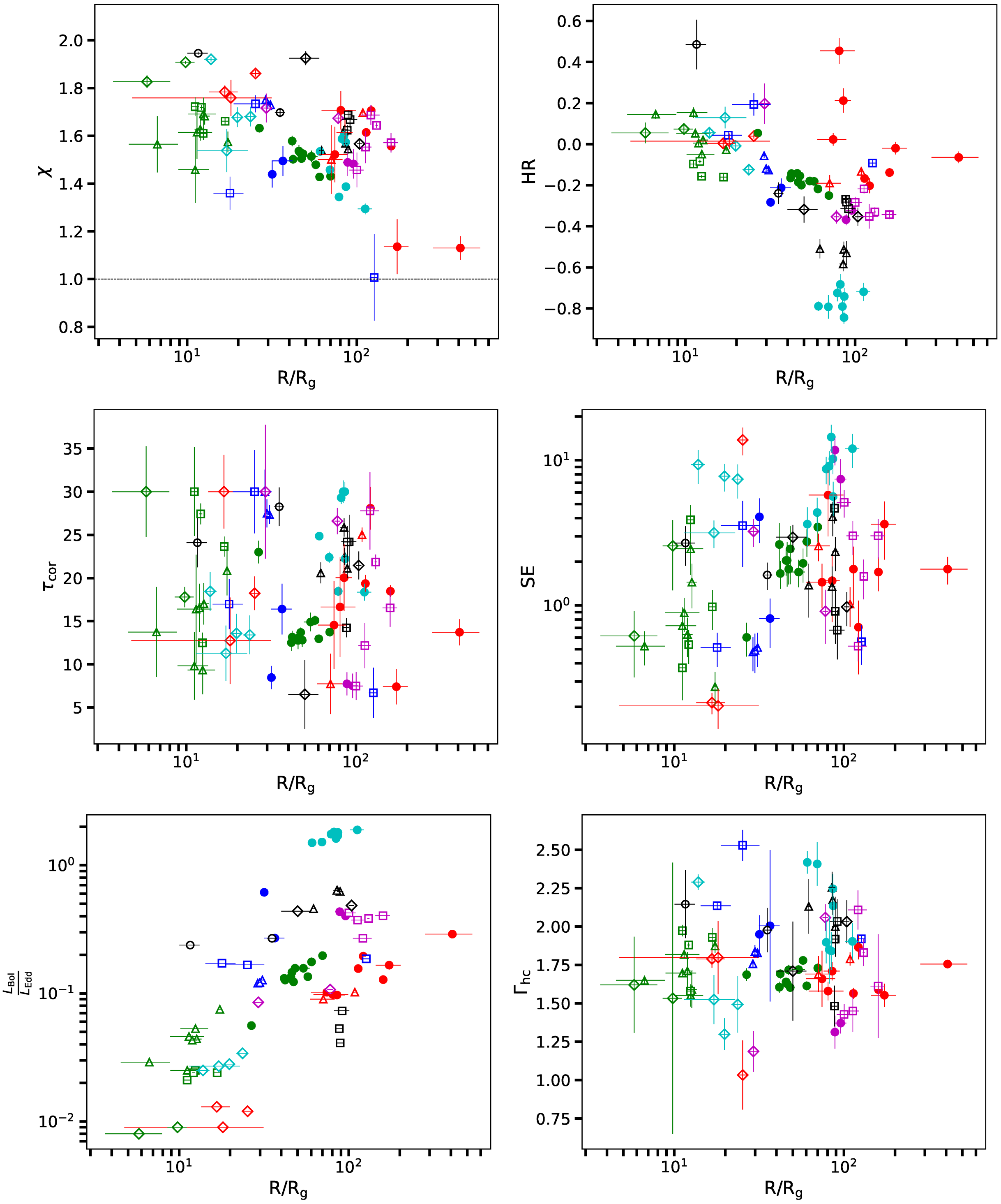

Fig. 10: Scatter plots showing the dependence of various warm corona properties and global parameters on the radius of warm corona, given in units of gravitational radius. There are a total 19 sources consisting of 78 observations; out of which, 8 sources have redshift-independent distance measurement, while for the remaining 11, the Hubble distance was used. See the main text for details. Unique markers indicate different sources, as in Fig. 6.

The black hole masses in our sample span two orders of magnitudes and the inner accretion flow may be at different evolutionary stages, as understood from hardness ratio plots presented in this paper. This leads to large uncertainties in the radius values and any tight correlation is difficult to obtain. Furthermore, the distance measurements were derived from two different methods which can also contribute to scatter. Nevertheless, we observe certain interesting trends in the data that point towards a transformation of the warm corona layer with changing disk-corona geometry.

The first trend is shown in the top-most left panel of Fig. 10, where the $\chi$ decreases as the warm corona expands, with a black dotted line separating the passive disk from the active disk. Similarly, on the top-most right panel, we observe that spectra be-come increasingly softer with expanding warm corona. This suggests the growing importance of soft X-ray excess emission with increasing size of warm corona. The cyan solid circles corresponding to RE1034+396 and red solid circles corresponding to 1H0419-577 add considerably to the dispersion. 1H0419-577 has complex absorption, which could lead to the hardening of spectra leading to large HR values. On the other hand, RE1034+396 is a super-Eddington accretor wherein, other physical processes might participate in soft X-ray excess production as discussed in Kaufman et al. (2017). Overall, comparing this trend to the variation of $\Gamma_{\rm hc}$ in the bottom right panel, we notice that it is independent of the warm corona radius. Similarly, $\tau_{\rm cor}$ seems to be independent of the size of the warm corona as well. This points towards the fact that while the extrinsic property of warm corona (such as size) drives the magnitude of soft X-ray excess emission relative to hard X-rays, the intrinsic properties of warm corona (such as $\tau_{\rm cor}$ and $\log q_{\rm h}$) are possibly coupled to variation of hot corona, as already demonstrated in Fig. 6. For instance, as in the middle left panel, we observe that $\tau_{\rm cor}$ does not show any coherent variation with a warm corona radius. The soft excess strength SE weakly depends on the warm corona radius, and it increases as the warm corona expands. This conclusion is weakened due to the presence of a large scatter which could be a result of SE depending on both the intrinsic and extrinsic properties of warm corona.

At the bottom left panel of Fig. 10, we present the connection of the accretion rate with the warm corona radius. An increase in $L_{\rm Bol}/L_{\rm Edd}$ is accompanied by the increase in warm corona radius, which again on close inspection, is obeyed by individual sources. This result is particularly interesting since the accretion rates were determined using distinct methods. It is well known that large changes in accretion rate are associated with changing inner disk structure. The source, Mrk 1018 underwent a large drop in mass accretion rate from 8% to 0.6% which was accompanied by 60% drop in the soft X-ray excess emission (Noda & Done 2018), Mrk 841 also displayed a diminishing optical/UV continuum, associated with a vanishing soft X-ray excess (Mehdipour et al. 2023) and similar conclusions were drawn for Mrk 590 as well (Denney et al. 2014). Our results demonstrate that such variation may be associated with the evolution of the warm corona which may depend on the global accretion rate and its intrinsic properties in a non-trivial way. Capturing the trigger of such evolution may be important in the context of "changing-state" AGNs, which exhibit correlated changes in the optical/UV continuum with accretion rate (Ricci & Trakhtenbrot 2023; Veronese et al. 2024, and references therein).

## 6. Discussion

### 6.1. Nature of the warm corona

We obtained an optically thick warm corona with mean value of $\tau_{\rm cor} = 18.26 \pm 0.12$, which agrees with vast majority of optical depth values previously reported (Petrucci et al. 2018; Jiang et al. 2019b; Middei et al. 2019, and references therein). Our results also indicate the requirement of relatively high internal heating of the warm corona, which shows two peaks. The first peak corresponds to sources with hard photon index for which $\log q_{\rm h} = -22.0$, and the second to sources with a soft and intermediate range of $\Gamma_{\rm hc}$ for which $\log q_{\rm h} = -22.5$. The mean internal heating for all observations results in $q_{\rm h} = 5.62\,(\pm 0.38) \times 10^{-23}$ erg s$^{-1}$ cm$^{3}$. Both results confirm the earlier expectation that the warm corona that is cooled by the Compton process must be internally heated to stay in balance with an accretion disk (Różańska et al. 2015).

Recently, a more realistic treatment of a magnetically heated disk and corona atmosphere confirmed the existence of a warm corona with an optical depth of up to 50 (Gronkiewicz et al. 2023). The value of the mean internal heating reported in this work belongs to a moderately high level of dissipation when compared to the parameter space considered in Gronkiewicz et al. (2023). Furthermore, the range of $\log q_{\rm h}$ resulting from our fitting procedure is prone to classical thermal instability, a key aspect of magnetically supported disks (MSDs) with the dominance of Comptonization and free–free emission, forming the warm corona.

All 82 observations, corresponding to 21 sources, point towards the existence of a dissipative warm corona on top of a cold, accretion disk. The emergent soft X-ray excess emission spectra are smooth, as is expected in a Compton-dominated warm layer with dissipation fraction ranging from 1 to 1.95. This is aligned with the claims of the copious amount of energy dissipation in the warm layer suggested by radiative transfer models (Ballantyne & Xiang 2020; Xiang et al. 2022; Kawanaka & Mineshige 2024) as well as magneto-hydrodynamic simulations (Hirose et al. 2006; Jiang et al. 2019b; Mishra et al. 2020), which could lead to the origin of such a warm layer.

We noticed a subtle correlation between the warm corona parameters with the photon index of the hot corona being modeled by the independent `NTHCOMP` model. It must be noted that we did not impose any a priori link between the parameters of the two Comptonizing coronae. All trends between model parameters are the result of their adjustment to the observed spectral shape. Our analysis indicates that hard sources (with lower value of $\Gamma_{\rm hc}$) are associated with warm corona having lower optical depth and higher internal heating. Conversely, softer sources indicate warm corona of higher optical depth and lower heating. It may be possible that at larger optical depths, the warm corona is acting as the source of extra seed photons entering the hot corona and cooling this region faster which results in softening the $\Gamma_{\rm hc}$ index. Such an instance was recently reported in the case of Mrk 359 (Middei et al. 2020), where the observed spectral variability detected by *XMM-Newton* and *NuSTAR* mission was interpreted by two-corona model, in which the outer disk is covered by a warm corona and the warm corona's photons may cool the hot corona through Comptonization. On the other hand, a flat $\Gamma_{\rm hc}$, indicating an extremely hard hot corona, can lead to the destruction of the warm corona by depriving it of seed photons and causing extreme heating.

The above ideas are model-dependent and subject to limitations for two reasons. First, we set strict constraints on the $\Gamma_{\rm hc}$ by freezing the temperature of the hot corona (Table 2). Secondly, we assumed in all `TITAN-NOAR` models the same photon index of 1.80 of the hard external illumination from the hot corona. While the above-mentioned value of photon index is a standard assumption for the majority of AGNs (Ricci et al. 2017; Akylas & Georgantopoulos 2021), it can affect the inferred warm corona properties, especially for AGNs with very steep and very flat $\Gamma_{\rm hc}$. Nevertheless, Petrucci et al. (2013) demonstrated that the hot corona illumination is a factor of two lower than the total warm corona luminosity, therefore, it has less of an effect on the warm corona properties than internal mechanical heating. In addition, Xiang et al. (2022), with the use of the `ReXcor` model, showed that the photon index of external illumination has a negligible effect on the emitted soft X-ray spectrum. We note that this was the case even for the low coronal height in their lamp-post model.

Table 3: Estimates of warm corona radius $R_{\rm cor}$ from `TITAN-NOAR` normalization $N_{\rm SE}$ (Eq. 6), in terms of gravitational radii, for sources of known distances ($D$).

| Source $D$ | ObsID | $N_{\rm SE} = S/D^2$ [$\times 10^{-7}$] | $R_{\rm cor}$ [$R_{\rm g}$] | Source $D$ | ObsID | $N_{\rm SE} = S/D^2$ [$\times 10^{-7}$] | $R_{\rm cor}$ [$R_{\rm g}$] |
|---|---|---|---|---|---|---|---|
| | 112600401 | 2.20±0.17 | 121.76 ± 4.58 | | 112170101 | 1.77 ±0.18 | 11.14 ±0.57 |
| | 148000201 | 0.97±0.45 | 80.62 ± 18.58 | NGC 7469 | 112170301 | 3.99 ±0.31 | 16.75 ± 0.65 |
| 1H0419-577 | 148000301 | 1.08±0.23 | 85.26 ± 9.24 | 57.30 Mpc | 207090101 | 2.20 ±0.16 | 12.45 ± 0.46 |
| | 148000401 | 24.80±15.0 | 408.55, ± 126.03 | | 207090201 | 2.10 ±0.08 | 12.15 ± 0.24 |
| 459.83 Mpc | 148000501 | 4.46±1.48 | 173.26 ± 28.75 | PG0804+761 | 605110101 | 1.47 ±0.23 | 95.72 ± 7.60 |
| | 148000601 | 0.82±0.32 | 74.28 ± 14.34 | 442.45 Mpc | 605110201 | 1.26 ±0.15 | 88.49 ± 5.27 |
| | 604720301 | 1.92±0.09 | 113.77 ± 2.79 | PG0844+349 | 103660201 | 1.25 ± 0.10 | 77.75 ± 2.99 |
| | 604720401 | 3.79±0.34 | 159.71 ± 7.10 | 390.00 Mpc | 554710101 | 0.18 ± 0.02 | 29.29 ± 1.98 |
| ESO198-G24 | 067190101 | 0.22 ± 0.08 | 16.67 ± 3.18 | | 201940101 | 0.88 ±0.02 | 130.86 ± 1.73 |
| 200.89 Mpc | 112910101 | 0.26 ± 0.38 | 18.12 ± 13.37 | PG1116+215 | 201940201 | 0.76 ±0.19 | 121.17 ± 15.20 |
| | 305370101 | 0.50 ± 0.05 | 25.31 ± 1.35 | | 554380101 | 1.30 ± 0.25 | 159.22 ± 15.48 |
| HE1029-1401 | 110950101 | 1.06± 0.35 | 70.79 ± 11.71 | 781.54 Mpc | 554380201 | 0.66 ±0.11 | 112.85 ± 9.35 |
| 384.82 Mpc | 203770101 | 2.51± 0.06 | 108.68 ± 1.28 | | 554380301 | 0.52 ±0.10 | 100.25 ± 9.54 |
| IRASF12397+3333 | 202180201 | 1.32±0.12 | 31.75 ± 1.47 | PG1351+640 | 205390301 | 0.26 ±0.03 | 35.30 ± 2.41 |
| 155.0 Mpc | 202180301 | 1.76±0.49 | 36.68 ± 5.07 | 391.23 Mpc | 556230201 | 0.03 ±0.01 | 11.62 ± 1.58 |
| MRK 279 | 302480401 | 1.64 ±0.08 | 30.96 ±0.77 | | | | |
| 135.49 Mpc | 302480501 | 1.45 ±0.06 | 29.06 ± 0.58 | PG1402+261 | 400200101 | 0.64 ±0.05 | 103.89 ± 3.76 |
| | 302480601 | 1.53 ±0.11 | 29.90 ± 1.04 | 728.43 Mpc | 400200201 | 0.15 ±0.06 | 49.99 ± 10.02 |
| MRK 335 | 510010701 | 2.72 ±1.39 | 25.28 ± 6.46 | | 5010101 | 1.87 ± 0.09 | 85.87 ± 2.00 |
| 85.90 Mpc | 600540501 | 68.5 ±8.01 | 126.84 ± 7.42 | PG1440+356 | 5010201 | 2.01 ±0.17 | 88.96 ± 3.69 |
| | 600540601 | 1.36 ±0.55 | 17.88 ± 3.59 | 351.87 Mpc | 5010301 | 0.98 ± 0.05 | 62.04 ± 1.70 |
| | | | | | 107660201 | 1.84 ±0.17 | 85.10 ± 3.94 |
| | 0130720101 | 6.92±0.52 | 26.69 ± 0.46 | | | | |
| | 0601390201 | 5.06±0.42 | 42.15 ± 1.76 | Q0056-363 | 102040701 | 0.50 ±0.10 | 91.38 ± 9.25 |
| | 0601390301 | 6.36±0.56 | 47.26 ± 2.10 | 722.24 Mpc | 205680101 | 0.47 ±0.03 | 87.98 ± 3.22 |
| | 0601390401 | 6.63±0.51 | 48.24 ± 1.85 | | 401930101 | 0.48 ± 0.04 | 88.99 ± 4.06 |
| Mrk 509 | 0601390501 | 5.01±0.16 | 60.41 ± 3.51 | | 109070101 | 4.03 ±0.22 | 69.57 ± 1.90 |
| | 0601390601 | 14.02±0.85 | 70.09 ± 2.12 | | 506440101 | 3.08 ±0.09 | 60.8 ± 0.91 |
| 105 Mpc | 0601390701 | 8.31±1.45 | 54.03 ± 4.73 | RE1034+396 | 561580201 | 6.17 ±0.33 | 86.05 ± 2.31 |
| | 0601390801 | 6.09±0.51 | 46.25 ± 1.95 | | 655310101 | 6.21 ±0.33 | 86.36 ±2.06 |
| | 0601390901 | 4.93±0.21 | 41.63 ± 0.91 | 194.16 Mpc | 655310201 | 5.15 ±0.18 | 78.62 ±1.40 |
| | 06013901001 | 9.38±0.72 | 57.38 ± 2.21 | | 675440101 | 10.5 ±1.99 | 112.25 ± 10.64 |
| | 06013901101 | 5.92±0.42 | 45.60 ± 1.95 | | 675440201 | 5.59 ±0.26 | 81.87 ± 1.93 |
| | | | | | 675440301 | 5.91 ±0.16 | 84.20 ± 1.16 |
| MRK 590 | 109130301 | 0.14 ±0.10 | 5.79 ± 2.13 | | | | |
| 87.10 Mpc | 201020201 | 0.40 ±0.10 | 9.79 ± 1.26 | | | | |
| | | | | | 103862101 | 1.14 ±0.76 | 17.12 ±5.69 |
| | 059830101 | 2.55 ±0.14 | 17.38 ± 0.41 | UGC393 | 400070201 | 2.17 ±0.36 | 23.64 ± 1.96 |
| | 109970101 | 1.31 ±0.49 | 12.48 ± 2.31 | | 400070301 | 0.74 ±0.13 | 13.82 ± 1.25 |
| NGC 4593 | 740920201 | 1.10 ±0.50 | 11.43 ± 2.31 | 90.00 Mpc | 400070401 | 1.52±0.24 | 19.77 ± 1.55 |
| | 740920301 | 1.05 ±0.43 | 11.15 ± 2.27 | | 502091001 | 1.00 ±0.15 | 16.06 ±1.22 |
| 61.00 Mpc | 740920401 | 0.38 ±0.24 | 6.67 ± 2.13 | | | | |
| | 740920501 | 1.36 ±0.19 | 12.69 ± 0.87 | | | | |
| | 740920601 | 1.20 ±0.17 | 11.94 ± 0.86 | | | | |

**Notes**:
Two sources- HB890405-123 and LBQS1228+1116 are not included here due to the unavailability of black hole mass measurements.
Details on distance measurement (D) are given in Sect. 5.4.

In the framework of our model, we did not observe the transition from the state of active warm corona and passive disk to a state where the disk would become dissipative. Gronkiewicz et al. (2023) demonstrated that classical thermal instability occurs at the base of the warm corona, and can trigger changes of disk-corona radiative equilibrium, leading to the build-up of the warm optically thick plasma above an accretion disk on the timescales of the order of days. Deeper multi-epoch data in different energy bands are needed to fully test this scenario.

One question remains to be answered regarding the origin of this heating inside the warm corona. The coincidence seen among the heating values within the realm of MSDs may give some hints that magnetic fields are responsible for heating upper layers of disk atmosphere (Hirose et al. 2006; Begelman et al. 2015; Gronkiewicz et al. 2023). Multi-dimensional MHD simulations have shown the requirement of strong magnetic fields at large scale height of stable accretion disks, is consistent with large optical depths $\gg 5$ of the warm plasma, as observed in our sample (Turner et al. 2003; Beckwith et al. 2009; Takeuchi et al. 2010; Mishra et al. 2020; Wielgus et al. 2022; Liska et al. 2022).

### 6.2. Comparison with the ReXcor model

Recently, a similar model based on the soft X-ray excess emission (i.e., ReXcor; (Ballantyne 2020; Xiang et al. 2022)) was tested on a smaller subset of sources from the same sample as in our work (Ballantyne et al. 2024, hereafter B24). The origin of the heating inside the warm corona in this model is connected to purely viscous dissipation of accretion energy by standard disk and the total dissipated energy is divided between three emitting regions: cold disk, warm corona, and hot corona. Given that a hot corona needs to be powered as well, the authors concluded that 50% of the total accretion energy powers the warm corona, which supports the existence of a large level of heating. Since, we did not restrict $\log q_{\rm h}$ to any particular physical mechanism, it provides an upper limit on a realistic level of heating that can feasibly exist in the warm corona. It would be interesting to explore heating levels for sources that possess active disks and we plan to study this in our future research.

Our implementation of the composite X-ray spectral model differs from B24 in two respects. While they approximated the hard X-ray tail of the spectra with a powerlaw, we used a more physical NTHCOMP model, self-consistently connecting the seed photon temperature of 7 eV. However, the main difference arises in the soft X-ray emission band where, we employ the TITAN-NOAR model, as opposed to ReXcor by B24. Overall, our total model has 3 degrees of freedom less than B24.

As a result of spectral fitting, the authors of B24 reached the same conclusion, namely, that the warm corona has to be highly dissipative, which indicates the dominance of external agents (e.g., the magnetic field) playing an important role in maintaining a warm corona. The authors concluded the presence of $\sim$ 50-70% heating fraction thus supporting the production of soft X-ray excess from warm corona. They obtained a mean optical depth of 14, which is close to our estimates. However, there is a considerable difference in the relation between optical depth and internal heating of the warm corona. In B24, the authors found a higher amount of heating fraction in warm corona for larger optical depth which is in opposition to our results. This may be an artifact of different model construction.

Next, B24 obtained a "v- shaped" trend between both warm corona optical depth and heating fraction with accretion rate. We did not observe any particular trend between the same parameters from our study. This could be attested to the fact that our model is essentially independent of an accretion rate. In fact, the total warm corona emission is mainly a result of reprocessed emission of the illuminated radiation originating in the hot corona as well as from the cold disk, along with dissipative emission from inside the warm corona. From our studies, we have merely observed the overall effect of how these micro-processes vary with global accretion rate which was estimated using a distinct method in P18. Thus, the overall differences in the setup of the two models as described above and in Sect. 2 lead to such trends. Nevertheless, they observed an increase in warm corona flux with observed accretion rates, which could be due to the increasing radial size of warm corona. We have drawn the same conclusion after estimating the size of warm corona and studying how it depends on accretion rates, thus strengthening the overall goals of both methods.

While the ReXcor model considers deposition of fraction of accretion energy in the form of reflection from lamppost-like corona, its effect is not very significant in determining the shape of soft X-ray excess (Ballantyne & Xiang 2020). The application of the ReXcor model to observations of a sample of Seyfert 1 galaxies revealed a low fraction of the total accretion energy budget contributing towards reflection (B24, Porquet et al. 2024), when compared to the amount of heating. Additionally, the lamppost geometry for hot corona is currently refuted by X-ray polarimetric measurements of a few AGNs (Pal et al. 2023; Tagliacozzo et al. 2023) and X-ray binaries (Jana & Chang 2024). Another difference between our model with ReXcor is that the latter includes the dependence on black hole spin. At high ionization levels, such as those considered in the warm corona, the effect of relativistic blurring is negligible. This effect has also been shown by Xiang et al. (2022) using ReXcor. Thus, our negligence of the spin has less of an effect on the inferred properties of the warm corona. Despite differences in the design of the two models, our global results are in agreement with the requirement of having extra heating inside warm corona to adequately describe the smooth, soft X-ray excess spectrum and increasing radial size of warm corona with the accretion rate.

### 6.3. Warm corona with an accretion disk

We found that the presence of warm corona in the inner regions of the accretion disk can adequately describe the soft X-ray excess feature in a variety of AGNs, spread across a wide range of accretion rates and black hole masses (Fig. 9). One of the most interesting results of our work is the evolution of the radial extent of warm corona with accretion rate (Fig. 10, bottom left). It must be noted that no link between Comptonized emission and soft UV emission was imposed a priori. The model adjusts its parameters to fit the data. We observed that as the accretion rate of the system increases, the warm corona radius increases. However, similar trends were not observed between accretion rate and warm corona properties or between warm corona radius and its properties. We speculate that this apparent disconnect could arise due to the standard disk playing a very different role from that of the hot corona in the formation and evolution of warm corona. Changes in the disk are more prone to driving the external properties of the warm layer such as its extent (Fig. 10, bottom-left panel) or partial contribution to soft X-ray excess emission (Fig. 9, bottom-right panel). On the other hand, innate features like heating and optical depth are largely influenced by the hot corona (hence, the overall inner disk geometry) as evident from trends shown in Fig. 6. A recent study by (Waddell et al. 2023), suggested external factors like winds or magnetic fields influencing the properties of warm corona. This study conducted on 200 AGNs in the eROSITA Final Equatorial Depth Survey (eFEDs) revealed an increase in soft X-ray excess emission with accretion rate. They associated a failed wind settling on the disk to be behind the formation of the warm corona. In such a scenario, the accretion rate will not directly influence warm corona optical depth or internal heating. It also depends at what location the winds settle since heating (as well as disk flux injecting into the warm corona) is not uniform throughout the radial extent of a disk.

In keeping with the prevailing idea of changing disk-corona system with accretion rate, at lower values of the accretion rate (typically a few percent of $L_{\rm Bol}/L_{\rm Edd}$, where the disk is thought to be truncated), we found that extension of the warm corona is lowest. This indirectly indicates that the surface area of the warm corona decreases as the disk recedes from the black hole, making the warm layer photon-starved. At such low $L_{\rm Bol}/L_{\rm Edd}$, hot flow can occur close to the black hole (Yuan & Narayan 2014) and any warm corona that has survived the absence of a cold disk now begins to disappear as it competes for seed photons against the powerful hot corona. A strong magnetic field powering the hot corona could leak into the warm corona as well, confining it to smaller optical depths. Such a modulation of the warm corona can play an important role in the changing-state behavior seen for a few AGNs (Noda et al. 2011; Mahmoud & Done 2020; Ricci & Trakhtenbrot 2023).

On the other hand, at $L_{\rm Bol}/L_{\rm Edd} > 0.1$, we noticed the increase in warm corona surface area, due to the appearance of the disk which replenishes the warm corona with seed photons for Compton cooling, thereby stabilizing it in the process. This is also reflected in slightly higher values of soft X-ray excess strength. In our work, it is interesting to notice how individual sources still obey an increasing trend even when they are located in different regions of accretion rate-warm corona radius scatter space, hinting at a variety of evolutionary stages for each source. In summary, we found that the expanse of the warm corona layer is intricately linked to the standard accretion disk and less dependent on the hot corona. We show that even with a crude estimate, changing accretion states in AGNs affect the warm corona and, hence, the emission in the soft X-ray band. This paves the way to testing such models or relations on multipoch and multiwavelength observations of singular sources, which are better equipped for understanding the changing complex inner-disk geometry.

## 7. Conclusions

In this work, we tested a model of dissipative warm corona on a sample of 21 AGNs with widely distributed redshifts, accretion rates, and black hole masses. The final emission responsible for soft X-ray excess was obtained by radiative transfer computations with the `TITAN-NOAR` code, where the internal heating was balanced by Compton cooling with all other radiative processes, such as photoionization and bremsstrahlung, were taken into account, simultaneously. The final grid of models for a wide range of parameters was used in the spectroscopic analysis of the 0.3–10 keV EPIC-pn data, allowing us to set constraints on the optical depth of the warm corona and the value of internal heating. Then, we searched for correlated trends of warm corona properties with global parameters of AGNs, such as accretion rate and hardness ratios. Finally, our model allowed us to estimate the extent of warm corona being in harmony with our assumed model of the inner disk geometry.

1. All observations point towards the existence of a dissipative warm corona with a dissipation fraction in the range between 1–1.95. Hence, most of the accretion energy is spent in the upper layers of warm corona. Changes in the dissipation fraction are not accompanied by substantial changes in hardness ratios, which suggests a common origin for the heating source for both the hot and warm corona.
2. The average optical depth of the warm corona in our sample is $\sim 18.26 \pm 0.12$ and the average internal heating $\sim 5.62(\pm0.38) \times 10^{-23}$ erg s$^{-1}$ cm$^{3}$. This region of parameter space is consistent with recent studies of the soft X-ray excess in AGNs. The emergent soft X-ray spectra are smooth, as expected due to Compton smearing of emission and absorption lines.
3. The sources with low and hard hot corona photon index are associated with assumption that the warm corona exhibits a lower optical depth, but stronger internal heating.
4. While the soft X-ray excess emission is common in sources spanning a wide range of accretion rates, the fundamental properties of the warm corona, such as the optical depth and internal heating, do not display any dependence on the accretion rate.
5. The radial expanse of warm corona varies through a large range of values, starting as low as 7 to 408 $R_{\rm g}$. For each source (and as a whole), we found a positive trend with accretion rate, suggesting a connection between warm corona and standard accretion disk.

Future works will improve on our warm corona emission model by tying together the parameters more tightly among the different model components used to build the total fitting model. On the other hand, combining deep broadband optical/UV data with hard X-rays will offer key insights into the disk-warm corona interplay.

*Acknowledgements.* This research was supported by the Polish National Science Center grants No. 2021/41/B/ST9/04110. BP would like to thank Alex Markowitz for riveting discussions on the spectral modeling of soft X-ray excess. BP would also like to thank colleague, Tathagata Saha for the helpful discussion on statistical & data analysis. This research was supported by the International Space Science Institute (ISSI) in Bern, through ISSI International Team project 514 (Warm Coronae in AGN: Observational Evidence and Physical Understanding). Part of this work has been financially supported by the National High Energy Program (PNHE) of the French National Center of Scientific Research (CNRS) and the French Space Agency (CNES). S.B. is an overseas researcher under the Postdoctoral Fellowship of Japan Society for the Promotion of Science (JSPS), supported by JSPS KAKENHI Grant Number JP23F23773. StB acknowledges funding from PRIN MUR 2022 SEAWIND 2022Y2T94C, supported by European Union - Next Generation EU, and from INAF LG 2023 BLOSSOM.

## Appendix A: Impact of a warm absorber

The warm ionized gas located along the line of sight towards the nucleus may absorb the soft X-ray continuum and therefore affect our analysis. The only way to eliminate the influence of eventual absorption is to implement a `CLOUDY` multiplicative model component as indicated by Eq. 7. We do it in our automatic fitting procedure, and also manually for several heavily absorbed sources, as described in Sect. 5.1. According to Fig. A.1, a significant number of sources show strong ionized host galaxy absorption with mean $\log N_{\rm H}^{\rm CL} = 21.23 \pm 0.76$ and $\log \xi_{\rm CL} = 2.02 \pm 0.68$. The best-fitted values of the warm absorber model can be found in Table D.3, where the second column is the column density and the third column is the ionization parameter. As mentioned in P18, the use of `CLOUDY` multiplicative table model is only to improve the fit statistic. Besides a few heavily obscured sources, it has an insignificant effect on other parameter values. Nevertheless, to test our results, we repeated the entire fitting process without the `CLOUDY` table model, which resulted in stronger residuals around 0.5-2 keV. However, trends among $\tau_{\rm cor}$, $q_{\rm h}$, and $\Gamma_{\rm hc}$ were approximately maintained, albeit with larger uncertainties.

As pointed out in Gierliński & Done (2004), complex absorption alone can reproduce the shape of soft X-ray excess, meaning that the interpretation of warm corona properties for heavily absorbed sources is not straightforward. Thus, a more accurate description of the complex absorption in these sources is necessary, requiring the construction of a new warm absorber model which is beyond the scope of this work.

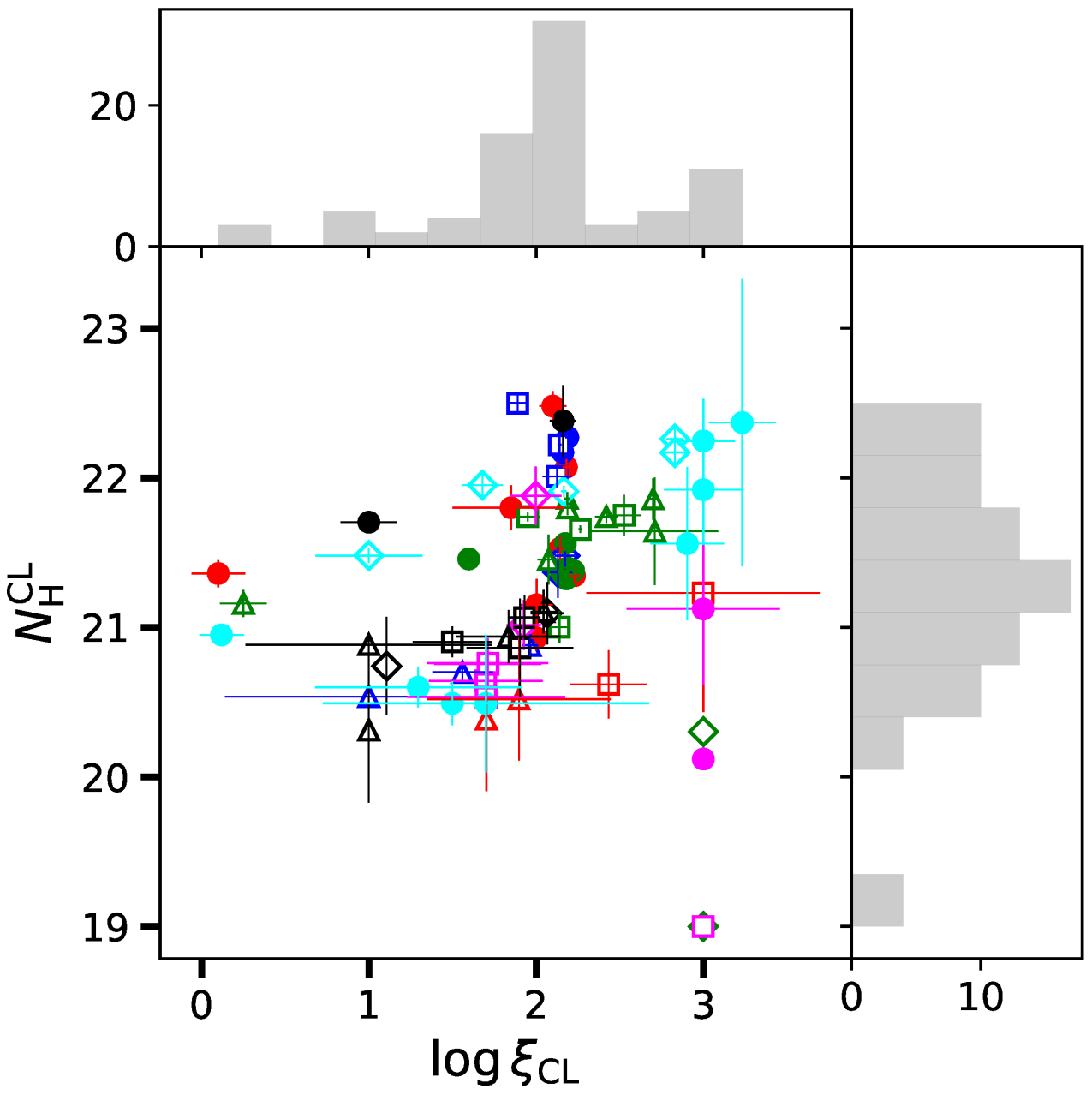

Fig. A.1: Scatter plot showing the best fitted hydrogen column density, $N_{\rm H}^{\rm CL}$, against ionization parameter, $\xi_{\rm CL}$, of `CLOUDY` model. The distribution of their values is shown as grey histograms. Different symbols associated with sources are the same as in Fig. 6.

## Appendix B: Pearson's correlations and their MCMC simulations

To assess the robustness of the strength of correlation estimated using Pearson's' method (Fig. 6), we performed MCMC simulations and obtained a distribution of correlation strength indicator denoted by r-value and the rejection probability denoted by p-value. We performed $10^6$ simulations by randomly sampling all the 82 data points within their error bars and calculated Pearson's r and p-value for all $10^6$ simulations. The results are shown in Fig. B.1. On the left panel, the distribution of the r-value is shown, which is characterized by a singly peaked approximate Gaussian shape. The mean, 5$^{\rm th}$ and 95$^{\rm th}$ percentiles are indicated by red vertical lines. On the right panel, their corresponding p-value distributions are shown. A p-value less than 0.05 is globally accepted as an indication of a high rejection probability of a null hypothesis or highly significant estimation of any quantity. We normalized the y-axis for all plots. Figure B.1 (left panel), showing the uni-modal distribution of r-values indicates weak correlation strength of the primary warm corona properties with $\Gamma_{\rm hc}$, but at a high significance level of $\gg 99\%$ (right panel).

## Appendix C: Impact of `RELXILL`- A prototypical case

Several AGNs have shown signatures of a broad iron K$\alpha$ line (equivalent width ~ 420-560 eV) and are best described by relativistic reflection models. One such model is `RELXILL`. It takes into account the effects of strong gravity on reflected emission close to the central black hole. It is often difficult to separate a broad component of the iron line from the continuum, which is a key requirement to infer physical properties such as the spin of the black hole ($a$), inner disk radius ($R_{\rm in}$) and emissivity. As shown by Guainazzi et al. (2006), about 200,000 counts in the hard X-ray band are necessary to ensure the presence of such broad features. To test relativistic effects on warm corona emission, we fit the *XMM-Newton* EPIC-pn spectra of a prototypical source-1H0416-577 which hosts broad iron lines and is well described by relativistic models (Jiang et al. 2019a). The source, 1H0419-577 displayed a broad Fe K$\alpha$ feature when the source was at a high flux state (Jiang et al. 2019a) in 2010 corresponding to ObsID: 0604720301. We replaced the `XILLVER` component with a basic flavor of `RELXILL` to study its effect on warm corona properties as well as inferring properties of the inner accretion disk.

We obtained a better fit with the new model, where $\Delta\chi^2$=15 for 5 extra degrees of freedom. During the fitting procedure, emissivity values were kept frozen at 5 and the break radius at 50 $R_{\rm g}$ since they could not be restricted. The high energy cutoff was fixed at 300 keV and the model output was set to reflected emission only. With an F-statistic value of 2.69 and p-value of 0.02, however, this is not a highly significant improvement over the original model described in Sect. 2. As shown in Table C.1, the `TITAN-NOAR` normalization reduced slightly due to the mild contribution of `RELXILL` to the soft X-ray excess (Fig. C.1). Comparing the best-fit results from Table D.2, there is a negligible impact seen for the $\tau_{\rm cor}$ and $\log q_{\rm h}$ of warm corona. Given the limited energy range, strong constraints on $R_{\rm in}$ and iron abundance ($A_{\rm fe}$) could not be obtained. In addition, `RELXILL` is a heavy model and has a large convergence time during the fitting process and MCMC runs. Thus, in the current scenario, `XILLVER` can adequately describe the EPIC-pn spectra of the majority of sources in our sample (by describing the reflected emission below 10 keV) at a lower computation time.

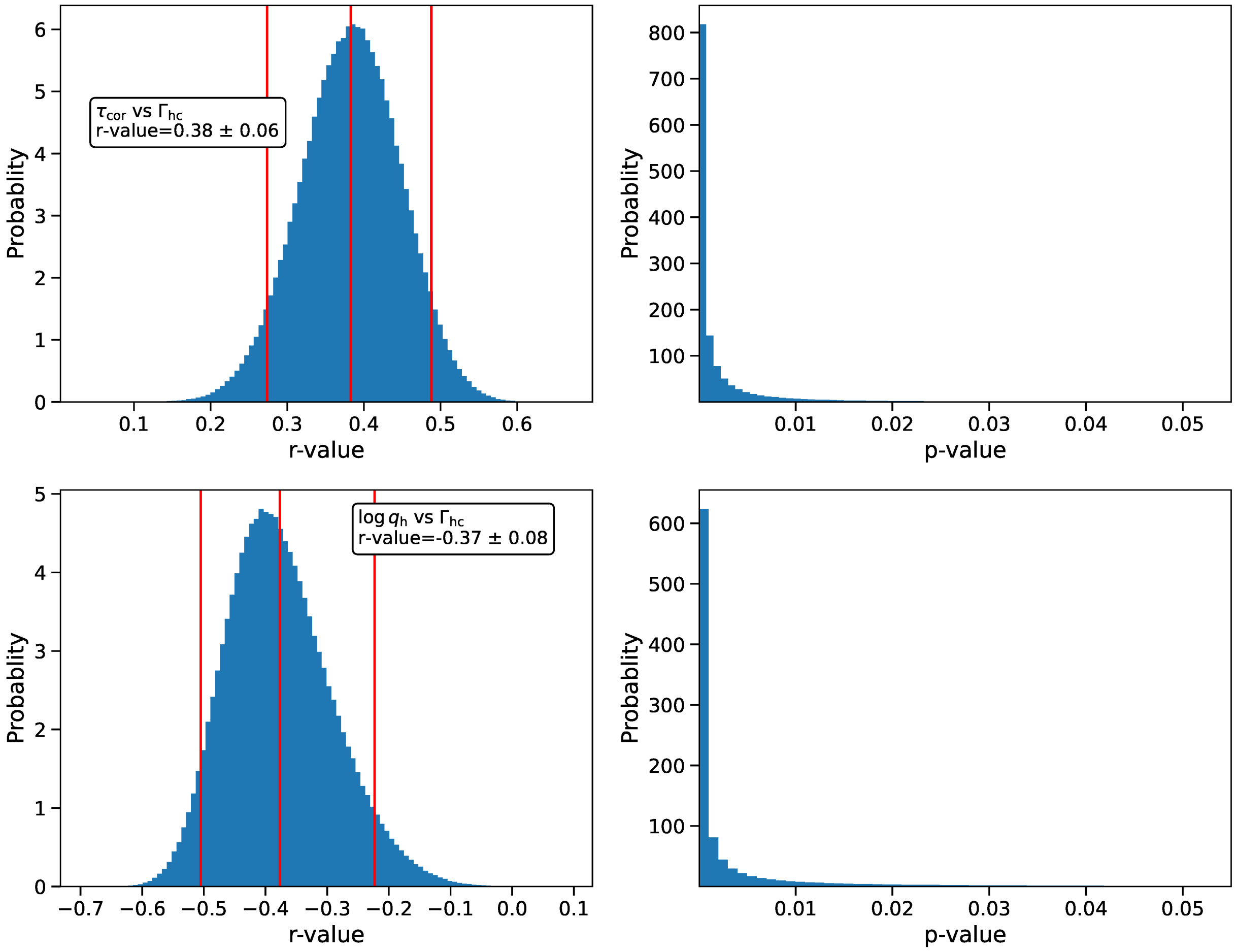

Fig. B.1: MCMC simulations showing the distribution of Pearson's r-value (left panel) and null hypothesis p-value (right panel) for warm corona optical depth $\tau_{cor}$ and log $q_h$ vs the $\Gamma_{hc}$.

Table C.1: All the best fitted parameters of the modified model as described in Appendix C applied to the source 1H0419-577 (ObsID: 0604720301).

| Model | Parameter | 1H0419-577 ObsID: 0604720301 |
|---|---|---|
| `WA` | log $\xi^{CL}$ | 1.89 ± 0.18 |
| | log $N_H^{CL}$ | 20.74 ± 0.13 |
| `zgauss` | E (keV) | 0.537 ± 0.005 |
| | $\sigma$ (eV) | 1.00* |
| | Norm ($\times 10^{-4}$) | 2.12 ± 0.52 |
| `nthComp` | $\Gamma_{hc}$ | 1.66 ± 0.11 |
| | Norm ($\times 10^{-3}$) | 2.68 ± 0.65 |
| `TITAN-NOAR` | $\tau_{cor}$ | 17.87 ± 1.49 |
| | log $q_h$ | -22.33 ± 0.09 |
| | Norm ($\times 10^{-7}$) | 1.75 ± 0.12 |
| `RELXILL` | $a$ | > 0.88 |
| | $R_{in}$ (in $R_g$) | -4.95 ± 6.50 |
| | $i$ (degrees) | 30.46 ± 5.55 |
| | $A_{fe}$ | 0.50 ± 0.45 |
| | log $\xi$ | 1.41 ± 0.38 |
| | Norm ($\times 10^{-5}$) | 7.22 ± 2.08 |

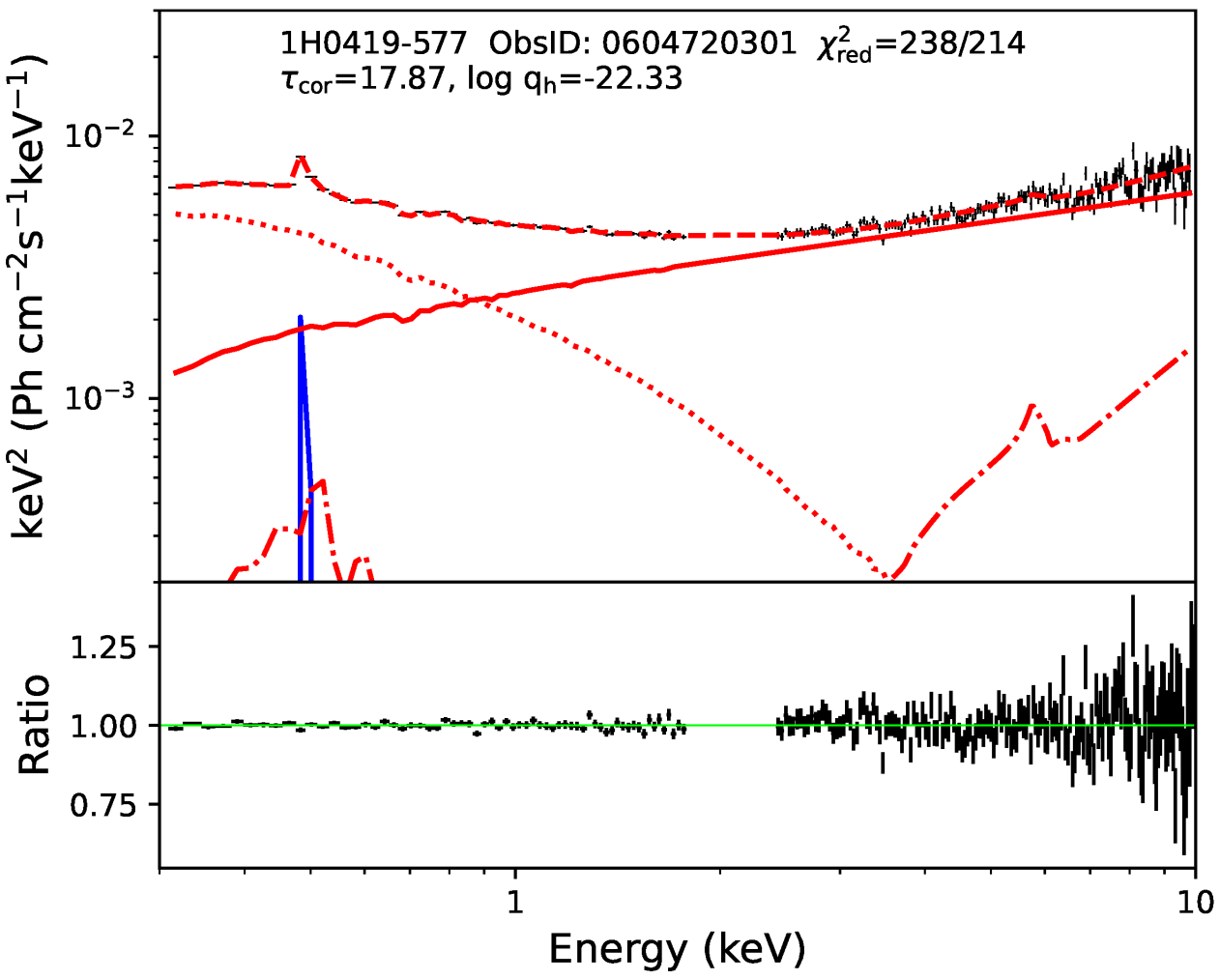

Fig. C.1: Example of an unfolded best-fit model of 1H0419-577 (ObsID: 0604720301) using relativistic reflection model and its data/model ratio, as described in Appendix C. The data are in black crosses. `RELXILL` model is represented by dashed-dotted line, `NTHCOMP` by solid line and `TITAN-NOAR` by a dotted line. A narrow Gaussian at 0.537 keV is shown as a blue solid line.

## Appendix D: Additional tables

Table D.1: All the best-fitted parameters for the ZPCFABS model used to fit the spectra of Mrk 335 described in Sect. 5.1

| | | ObsID | | |
|---|---|---|---|---|
| Model | Parameter | 0510010701 | 0600540501 | 0600540601 |
| zpcfabs | $\log N_H$ | 23.05 ± 0.87 | 22.69 ± 0.52 | 19* |
| | $C_{cov}$ (%) | 90.64 ± 0.03 | 25.30 ± 0.02 | 95* |
| XILLVER | $\log \xi_X$ | 0.00* | 1.21 ± 0.05 | 1.33 ± 0.07 |

**Note**:
'*' denotes that the value of the parameter was unconstrained, hence it was kept frozen.

Table D.2: Physically important quantities associated with our sample.

| Source ObsID | $\chi^2_{red}$ | $\frac{L_{Bol}}{L_{Edd}}$ | HR | $\Gamma_{hc}$ | $\log q_h$ | $\tau_{cor}$ | $\chi$ | SE | $[\times 10^{-11}]$ $F_{0.3-10}$ |
|---|---|---|---|---|---|---|---|---|---|
| 1H0419-577 | | | | | | | | | |
| 112600401 | 0.94 | 0.195 | -0.20 ± 0.04 | 1.87 ± 0.08 | -22.51 ± 0.06 | 28.09 ± 2.48 | 1.71 ± 0.02 | 0.70 ± 0.37 | 3.85 ± 0.27 |
| 148000201 | 1.32 | 0.098 | 0.45 ± 0.06 | 1.70 ± 0.06 | -22.07 ± 0.38 | 16.65 ± 5.77 | 1.71 ± 0.07 | 5.75 ± 2.75 | 1.05 ± 0.07 |
| 148000301 | 1.30 | 0.097 | 0.21 ± 0.06 | 1.71 ± 0.06 | -22.37 ± 0.21 | 20.04 ± 3.87 | 1.61 ± 0.05 | 1.47 ± 0.71 | 1.27 ± 0.09 |
| 148000401 | 1.03 | 0.290 | -0.06 ± 0.03 | 1.76 ± 0.02 | -22.56 ± 0.05 | 13.72 ± 1.52 | 1.13 ± 0.05 | 1.77 ± 0.22 | 2.34 ± 0.11 |
| 148000501 | 1.19 | 0.166 | -0.02 ± 0.03 | 1.55 ± 0.08 | -22.05 ± 0.21 | 7.42 ± 2.04 | 1.14 ± 0.11 | 3.63 ± 1.55 | 2.03 ± 0.14 |
| 148000601 | 1.08 | 0.102 | 0.02 ± 0.03 | 1.66 ± 0.18 | -22.22 ± 0.41 | 14.55 ± 5.07 | 1.52 ± 0.12 | 1.45 ± 0.45 | 1.89 ± 0.13 |
| 604720301 | 1.15 | 0.156 | -0.17 ± 0.01 | 1.56 ± 0.04 | -22.34 ± 0.02 | 19.35 ± 1.51 | 1.61 ± 0.01 | 1.65 ± 0.71 | 3.16 ± 0.07 |
| 604720401 | 1.15 | 0.128 | -0.14 ± 0.01 | 1.59 ± 0.04 | -22.38 ± 0.03 | 18.47 ± 0.69 | 1.56 ± 0.01 | 1.69 ± 0.44 | 2.91 ± 0.14 |
| ESO198-G24 | | | | | | | | | |
| 067190101 | 1.29 | 0.013 | 0.004 ± 0.014 | 1.79 ± 0.06 | -22.41 ± 0.20 | 30.00 ± 13.25 | 1.78 ± 0.08 | 0.21 ± 0.03 | 2.55 ± 0.01 |
| 112910101 | 1.20 | 0.009 | 0.01 ± 0.04 | 1.80 ± 0.24 | -21.75 ± 0.55 | 12.75 ± 10.00 | 1.76 ± 0.15 | 0.20 ± 0.06 | 1.98 ± 0.05 |
| 305370101 | 1.11 | 0.012 | 0.04 ± 0.01 | 1.03 ± 0.23 | -21.79 ± 0.10 | 18.24 ± 1.95 | 1.86 ± 0.01 | 13.83 ± 2.96 | 1.949 ± 0.005 |
| HB890405-123 | | | | | | | | | |
| 202210301 | 0.90 | - | -0.09 ± 0.02 | 1.78 ± 0.07 | -22.34 ± 0.08 | 16.46 ± 1.54 | 1.51 ± 0.03 | 0.67 ± 0.17 | 0.952 ± 0.003 |
| 202210401 | 1.00 | - | -0.08 ± 0.02 | 1.66 ± 0.05 | -21.92 ± 0.14 | 9.31 ± 1.66 | 1.46 ± 0.06 | 1.22 ± 0.37 | 0.975 ± 0.007 |
| HE1029-1401 | | | | | | | | | |
| 110950101 | 1.12 | 0.090 | -0.19 ± 0.04 | 1.69 ± 0.12 | -21.73 ± 0.37 | 7.75 ± 3.49 | 1.50 ± 0.14 | 2.57 ± 0.55 | 3.25 ± 0.15 |
| 203770101 | 1.16 | 0.102 | -0.13 ± 0.01 | 1.79 ± 0.04 | -22.43 ± 0.03 | 25.02 ± 0.82 | 1.70 ± 0.01 | 1.01 ± 0.31 | 4.85 ± 0.01 |
| IRASF12397+3333 | | | | | | | | | |
| 202180201 | 1.21 | 0.615 | -0.28 ± 0.02 | 1.95 ± 0.12 | -21.87 ± 0.13 | 8.48 ± 1.34 | 1.44 ± 0.06 | 4.07 ± 1.38 | 1.301 ± 0.001 |
| 202180301 | 1.00 | 0.270 | -0.21 ± 0.05 | 2.01 ± 0.49 | -22.35 ± 0.23 | 16.41 ± 2.93 | 1.49 ± 0.06 | 0.81 ± 0.30 | 1.04 ± 0.05 |
| LBQS1228+1116 | | | | | | | | | |
| 306630101 | 0.88 | - | -0.22 ± 0.04 | 1.81 ± 0.12 | -22.43 ± 0.06 | 30.00 ± 6.42 | 1.78 ± 0.04 | 1.08 ± 0.28 | 0.28 ± 0.01 |
| 306630201 | 0.97 | - | -0.22 ± 0.03 | 1.84 ± 0.10 | -22.42 ± 0.06 | 30.00 ± 5.69 | 1.78 ± 0.04 | 1.01 ± 0.33 | 0.30 ± 0.01 |
| MRK279 | | | | | | | | | |
| 302480401 | 1.23 | 0.127 | -0.13 ± 0.01 | 1.83 ± 0.03 | -22.45 ± 0.04 | 27.34 ± 1.08 | 1.73 ± 0.01 | 0.51 ± 0.13 | 6.04 ± 0.00 |
| 302480501 | 1.28 | 0.120 | -0.06 ± 0.01 | 1.76 ± 0.02 | -22.48 ± 0.03 | 30.00 ± 2.54 | 1.75 ± 0.02 | 0.48 ± 0.12 | 5.57 ± 0.39 |
| 302480601 | 0.98 | 0.121 | -0.12 ± 0.01 | 1.84 ± 0.04 | -22.46 ± 0.05 | 27.49 ± 1.63 | 1.73 ± 0.01 | 0.49 ± 0.14 | 5.59 ± 0.26 |
| MRK335 | | | | | | | | | |
| 510010701 | 1.25 | 0.167 | 0.19 ± 0.05 | 2.53 ± 0.10 | -22.52 ± 0.09 | 30.0 ± 10.81 | 1.73 ± 0.08 | 3.58 ± 1.70 | 0.61 ± 0.04 |
| 600540501 | 1.21 | 0.186 | -0.09 ± 0.02 | 1.92 ± 0.03 | -22.08 ± 0.05 | 6.70 ± 2.49 | 1.00 ± 0.18 | 0.63 ± 0.19 | 1.28 ± 0.03 |
| 600540601 | 1.60 | 0.172 | 0.04 ± 0.02 | 1.98 ± 0.02 | -22.52 ± 0.03 | 16.97 ± 2.83 | 1.36 ± 0.03 | 0.53 ± 0.13 | 1.00 ± 0.02 |
| MRK509 | | | | | | | | | |
| 130720101 | 1.24 | 0.056 | 0.05 ± 0.01 | 1.69 ± 0.04 | -22.46 ± 0.05 | 23.00 ± 1.30 | 1.63 ± 0.02 | 0.62 ± 0.15 | 5.92 ± 0.14 |
| 601390201 | 1.17 | 0.127 | -0.14 ± 0.01 | 1.69 ± 0.03 | -22.16 ± 0.06 | 13.16 ± 0.73 | 1.50 ± 0.02 | 1.69 ± 0.35 | 10.80 ± 0.25 |
| 601390301 | 1.29 | 0.123 | -0.15 ± 0.01 | 1.72 ± 0.03 | -22.19 ± 0.08 | 13.70 ± 0.57 | 1.50 ± 0.01 | 1.77 ± 0.38 | 11.26 ± 0.26 |
| 601390401 | 1.29 | 0.157 | -0.20 ± 0.01 | 1.60 ± 0.02 | -22.11 ± 0.08 | 12.81 ± 0.77 | 1.52 ± 0.02 | 2.46 ± 1.05 | 13.41 ± 0.93 |
| 601390501 | 1.36 | 0.176 | -0.22 ± 0.01 | 1.61 ± 0.03 | -22.23 ± 0.03 | 12.97 ± 0.43 | 1.42 ± 0.02 | 2.65 ± 0.59 | 11.29 ± 0.26 |
| 601390601 | 1.25 | 0.197 | -0.25 ± 0.01 | 1.73 ± 0.03 | -22.28 ± 0.03 | 13.74 ± 0.46 | 1.43 ± 0.01 | 3.47 ± 0.74 | 13.84 ±0.65 |
| 601390701 | 1.26 | 0.157 | -0.18 ± 0.01 | 1.72 ± 0.02 | -22.25 ± 0.11 | 14.91 ± 1.08 | 1.51 ± 0.02 | 1.69 ± 0.36 | 13.61 ± 0.32 |
| 601390801 | 1.21 | 0.146 | -0.19 ± 0.01 | 1.63 ± 0.02 | -22.12 ± 0.07 | 13.14 ± 0.66 | 1.53 ± 0.02 | 2.05 ± 0.69 | 13.01 ± 0.30 |
| 601390901 | 1.26 | 0.131 | -0.16 ± 0.01 | 1.61 ± 0.04 | -22.03 ± 0.07 | 12.52 ± 0.92 | 1.58 ± 0.02 | 2.65 ± 1.05 | 13.49 ± 0.31 |
| 601391001 | 1.24 | 0.135 | -0.18 ± 0.01 | 1.78 ± 0.02 | -22.3 ± 0.04 | 15.09 ± 0.55 | 1.48 ± 0.01 | 1.69 ± 0.44 | 12.24 ± 1.42 |
| 601391101 | 1.18 | 0.134 | -0.14 ± 0.01 | 1.64 ± 0.03 | -22.10 ± 0.08 | 12.78 ± 1.00 | 1.54 ± 0.03 | 2.04 ± 0.43 | 12.86 ± 0.30 |
| MRK590 | | | | | | | | | |
| 109130301 | 0.87 | 0.008 | 0.05 ± 0.05 | 1.62 ± 0.31 | -22.31 ± 0.36 | 30.0 ± 22.25 | 1.83 ± 0.11 | 0.62 ± 0.29 | 0.90 ± 0.06 |
| 201020201 | 1.12 | 0.009 | 0.07 ± 0.02 | 1.53 ± 0.88 | -21.57 ± 0.06 | 17.8 ± 1.17 | 1.91 ± 0.01 | 2.57 ± 0.29 | 1.23 ± 0.03 |
| NGC4593 | | | | | | | | | |
| 059830101 | 1.49 | 0.075 | -0.03 ± 0.01 | 1.87 ± 0.04 | -22.45 ± 0.04 | 20.84 ± 2.85 | 1.57 ± 0.01 | 0.45 ± 0.12 | 8.18 ± 0.19 |
| 109970101 | 1.35 | 0.053 | -0.05 ± 0.02 | 1.55 ± 0.07 | -21.62 ± 0.27 | 9.35 ± 2.80 | 1.69 ± 0.07 | 2.45 ± 0.83 | 8.21 ± 0.19 |
| 740920201 | 1.42 | 0.056 | 0.05 ± 0.02 | 1.82 ± 0.03 | -22.20 ± 0.07 | 16.50 ± 6.56 | 1.61 ± 0.04 | 0.89 ± 0.24 | 4.25 ± 0.10 |
| 740920301 | 1.10 | 0.025 | 0.15 ± 0.03 | 1.70 ± 0.02 | -21.97 ± 0.05 | 9.84 ± 3.92 | 1.45 ± 0.04 | 0.72 ± 0.24 | 2.35 ± 0.11 |
| 740920401 | 1.47 | 0.029 | 0.15 ± 0.02 | 1.65 ± 0.03 | -22.12 ± 0.79 | 13.75 ± 10.21 | 1.56 ± 0.23 | 0.52 ± 0.13 | 2.54 ± 0.06 |
| 740920501 | 1.20 | 0.044 | 0.02 ± 0.01 | 1.59 ± 0.12 | -22.13 ± 0.23 | 17.02 ± 2.39 | 1.68 ± 0.03 | 1.44 ± 0.49 | 5.30 ± 0.61 |
| 740920601 | 1.28 | 0.043 | 0.01 ± 0.01 | 1.71 ± 0.12 | -22.19 ± 0.23 | 16.55 ± 2.78 | 1.63 ± 0.05 | 0.63 ± 0.19 | 5.12 ± 0.83 |
| NGC7469 | | | | | | | | | |
| 112170101 | 1.17 | 0.021 | -0.10 ± 0.02 | 1.97 ± 0.04 | -22.54 ± 0.06 | 30.00 ± 5.13 | 1.72 ± 0.04 | 0.38 ± 0.15 | 6.01 ± 0.14 |
| 112170301 | 1.06 | 0.024 | -0.16 ± 0.02 | 2.01 ± 0.02 | -22.44 ± 0.06 | 23.65 ± 1.15 | 1.66 ± 0.01 | 0.97 ± 0.15 | 7.05 ± 1.15 |
| 207090101 | 1.33 | 0.025 | -0.16 ± 0.01 | 1.58 ± 0.02 | -21.98 ± 0.08 | 12.50 ± 1.19 | 1.61 ± 0.03 | 3.89 ± 1.02 | 7.33 ± 0.67 |
| 207090201 | 1.21 | 0.024 | -0.08 ± 0.01 | 1.88 ± 0.02 | -22.47 ± 0.04 | 27.43 ± 1.20 | 1.72 ± 0.01 | 0.53 ± 0.14 | 6.81 ± 0.62 |
| PG0804+761 | | | | | | | | | |
| 605110101 | 0.97 | 0.402 | -0.32 ± 0.02 | 1.37 ± 0.07 | -21.73 ± 0.15 | 7.52 ± 1.41 | 1.48 ± 0.06 | 7.41 ± 2.75 | 3.23 ± 0.01 |
| 605110201 | 1.12 | 0.434 | -0.37 ± 0.03 | 1.31 ± 0.11 | -21.74 ± 0.13 | 7.75 ± 1.34 | 1.49 ± 0.06 | 11.63 ± 2.52 | 2.51 ± 0.06 |
| PG0844+349 | | | | | | | | | |

Table D.2: continued.

| Source ObsID | $\chi^2_{red}$ | $\frac{L_{Bol}}{L_{Edd}}$ | HR | $\Gamma_{hc}$ | $\log q_h$ | $\tau_{cor}$ | $\chi$ | SE | [$\times10^{-11}$] $F_{0.3-10}$ |
|---|---|---|---|---|---|---|---|---|---|
| 103660201 | 1.16 | 0.107 | -0.35 ± 0.03 | 2.06 ± 0.09 | -22.52 ± 0.04 | 26.60 ± 1.52 | 1.67 ± 0.01 | 0.91 ± 0.36 | 1.64 ± 0.04 |
| 554710101 | 1.05 | 0.085 | 0.20 ± 0.10 | 1.19 ± 0.13 | -22.55 ± 0.05 | 30.00 ± 7.75 | 1.72 ± 0.06 | 3.23 ± 0.69 | 0.193 ± 0.001 |
| PG1116+215 | | | | | | | | | |
| 201940101 | 1.02 | 0.384 | -0.33 ± 0.02 | 1.83 ± 0.09 | -22.4 ± 0.04 | 21.86 ± 0.85 | 1.64 ± 0.01 | 1.58 ± 0.48 | 0.991 ± 0.002 |
| 201940201 | 1.04 | 0.392 | -0.35 ± 0.06 | 2.11 ± 0.13 | -22.54 ± 0.10 | 27.78 ± 4.47 | 1.69 ± 0.04 | 0.53 ± 0.11 | 1.06 ± 0.01 |
| 554380101 | 1.06 | 0.404 | -0.34 ± 0.02 | 1.61 ± 0.34 | -22.27 ± 0.18 | 16.55 ± 2.18 | 1.57 ± 0.04 | 3.04 ± 0.91 | 1.42 ± 0.07 |
| 554380201 | 1.16 | 0.373 | -0.22 ± 0.02 | 1.45 ± 0.14 | -22.03 ± 0.21 | 12.18 ± 2.62 | 1.55 ± 0.07 | 3.01 ± 0.79 | 1.12 ± 0.05 |
| 554380301 | 1.16 | 0.425 | -0.28 ± 0.03 | 1.43 ± 0.07 | -21.75 ± 0.17 | 7.50 ± 1.62 | 1.46 ± 0.07 | 5.01 ± 1.10 | 0.91 ± 0.02 |
| PG1351+640 | | | | | | | | | |
| 205390301 | 1.08 | 0.269 | -0.24 ± 0.05 | 1.98 ± 0.14 | -22.53 ± 0.05 | 28.26 ± 2.22 | 1.70 ± 0.02 | 1.64 ± 0.34 | 0.16 ± 0.01 |
| 556230201 | 1.07 | 0.238 | 0.49 ± 0.12 | 2.15 ± 0.22 | -21.59 ± 0.29 | 24.1 ± 15.28 | 1.95 ± 0.03 | 2.68 ± 0.81 | 0.067 ± 0.004 |
| PG1402+261 | | | | | | | | | |
| 400200101 | 0.99 | 0.485 | -0.35 ± 0.04 | 2.03 ± 0.14 | -22.49 ± 0.07 | 21.46 ± 1.60 | 1.57 ± 0.02 | 0.97 ± 0.25 | 0.52 ± 0.01 |
| 400200201 | 1.09 | 0.438 | -0.32 ± 0.06 | 1.71 ± 0.22 | -21.59 ± 0.42 | 6.52 ± 3.97 | 1.93 ± 0.03 | 0.32 ± 0.06 | 0.39 ± 0.01 |
| PG1440+356 | | | | | | | | | |
| 005010101 | 1.27 | 0.635 | -0.51 ± 0.04 | 2.17 ± 0.10 | -22.53 ± 0.04 | 22.50 ± 0.99 | 1.57 ± 0.01 | 4.07 ± 1.07 | 1.36 ± 0.03 |
| 005010201 | 1.01 | 0.625 | -0.53 ± 0.06 | 2.00 ± 0.19 | -22.51 ± 0.05 | 21.13 ± 1.15 | 1.54 ± 0.02 | 2.34 ± 0.61 | 1.18 ± 0.05 |
| 005010301 | 1.10 | 0.458 | -0.51 ± 0.05 | 2.13 ± 0.18 | -22.49 ± 0.06 | 20.59 ± 1.20 | 1.54 ± 0.02 | 1.40 ± 0.55 | 0.72 ± 0.02 |
| 107660201 | 0.76 | 0.635 | -0.58 ± 0.04 | 2.26 ± 0.10 | -22.56 ± 0.03 | 25.86 ± 1.05 | 1.63 ± 0.01 | 1.32 ± 0.50 | 1.26 ± 0.05 |
| Q0056-363 | | | | | | | | | |
| 102040701 | 0.84 | 0.073 | -0.31 ± 0.04 | 2.03 ± 0.14 | -22.45 ± 0.10 | 24.20 ± 3.11 | 1.67 ± 0.03 | 0.68 ± 0.25 | 0.80 ± 0.02 |
| 205680101 | 1.31 | 0.053 | -0.27 ± 0.02 | 1.48 ± 0.13 | -22.07 ± 0.10 | 14.24 ± 1.11 | 1.62 ± 0.02 | 4.69 ± 1.00 | 0.790 ± 0.002 |
| 401930101 | 0.98 | 0.041 | -0.28 ± 0.03 | 1.92 ± 0.12 | -22.41 ± 0.07 | 24.18 ± 1.83 | 1.69 ± 0.02 | 0.91 ± 0.36 | 0.74 ± 0.02 |
| RE1034+396 | | | | | | | | | |
| 109070101 | 1.25 | 1.519 | -0.79 ± 0.06 | 2.41 ± 0.14 | -22.65 ± 0.01 | 22.40 ± 0.64 | 1.46 ± 0.01 | 4.34 ± 1.47 | 0.97 ±0.02 |
| 506440101 | 1.11 | 1.500 | -0.79 ± 0.02 | 2.42 ± 0.07 | -22.65 ± 0.01 | 24.85 ± 0.42 | 1.53 ± 0.01 | 3.60 ± 1.10 | 0.976 ± 0.002 |
| 561580201 | 1.49 | 1.698 | -0.84 ± 0.03 | 2.25 ± 0.09 | -22.77 ± 0.02 | 30.00 ± 1.10 | 1.57 ± 0.01 | 10.22 ± 2.19 | 1.404 ± 0.001 |
| 655310101 | 1.12 | 1.804 | -0.74 ± 0.04 | 2.14 ± 0.10 | -22.72 ± 0.01 | 22.23 ± 0.76 | 1.39 ± 0.01 | 5.67 ± 1.47 | 1.05 ± 0.02 |
| 655310201 | 1.30 | 1.750 | -0.72 ± 0.04 | 1.90 ± 0.14 | -22.61 ± 0.01 | 18.46 ± 0.56 | 1.34 ± 0.01 | 8.71 ± 2.39 | 0.96 ± 0.02 |
| 675440101 | 1.23 | 1.889 | -0.72 ± 0.04 | 1.90 ± 0.12 | -22.65 ± 0.01 | 18.36 ± 0.98 | 1.29 ± 0.02 | 12.06 ± 3.10 | 1.09 ± 0.03 |
| 675440201 | 0.84 | 1.819 | -0.68 ± 0.05 | 1.85 ± 0.22 | -22.63 ± 0.01 | 19.3 ± 0.71 | 1.36 ± 0.01 | 9.12 ± 2.39 | 1.05 ± 0.02 |
| 675440301 | 1.17 | 1. 691 | -0.79 ± 0.04 | 1.50 ± 0.12 | -22.73 ± 0.01 | 29.98 ± 1.30 | 1.60 ± 0.01 | 14.45 ± 3.10 | 1.66 ± 0.04 |
| UGC3973 | | | | | | | | | |
| 103862101 | 1.32 | 0.027 | 0.13 ± 0.05 | 1.52 ± 0.16 | -21.99 ± 0.59 | 11.29 ± 8.20 | 1.54 ± 0.23 | 3.16 ± 0.67 | 2.30 ± 0.16 |
| 400070201 | 1.13 | 0.034 | -0.12 ± 0.02 | 1.49 ± 0.18 | -21.93 ± 0.15 | 13.42 ± 2.24 | 1.68 ± 0.04 | 7.41 ± 1.94 | 5.68 ± 0.13 |
| 400070301 | 1.19 | 0.025 | 0.06 ± 0.02 | 2.29 ± 0.05 | -21.54 ± 0.12 | 18.46 ± 2.25 | 1.92 ± 0.01 | 9.33 ± 2.45 | 4.07 ± 0.09 |
| 400070401 | 1.21 | 0.028 | -0.01 ± 0.02 | 1.30 ± 0.10 | -21.95 ± 0.17 | 13.60 ± 2.26 | 1.68 ± 0.04 | 8.15 ± 1.74 | 4.16 ± 0.10 |
| 502091001 | 1.43 | 0.025 | 0.42 ± 0.05 | 1.86 ± 0.04 | -22.31 ± 0.11 | 22.00 ± 5.08 | 1.70 ± 0.06 | 0.57 ± 0.17 | 1.06 ± 0.02 |

**Notes:**
The best fitted parameters- hot corona photon index ($\Gamma_{hc}$), total internal heating ($\log q_h$) and optical depth of warm corona ($\tau_{cor}$) of the model- `TBABS*CLOUDY*(TITAN/NOAR+NTHCOMP+XILLVER)` as described in Sect. 2 are presented in columns 5-7 respectively.
Goodness of fit indicator- reduced Chi-square values ($\chi^2_{ref}$) are listed in second column.
Values of accretion rate ($L_{Bol}/L_{Edd}$) adopted from P18, are mentioned in the third column, where '-' denotes missing values.
Hardness ratios (HR) were computed using count rates are listed in fourth column.
Dissipation rate ($\chi$), were calculated a posteriori using Eqn 4 mentioned in eighth column.
Soft excess strength (SE) were calculated using Eq. 9 mentioned in ninth column.
$F_{0.3-10}$ is the un-absorbed flux in the energy range 0.3–10 keV, expressed in units of ergs $s^{-1}$ $cm^{-2}$, mentioned in tenth column.

Table D.3: Additional best fitted parameters for all objects in our sample

| Source<br>ObsID | $\log N_H^{CL}$ | $\log \xi_{CL}$ | [$\times 10^{-3}$]<br>$N_{hc}$ | [$\times 10^{-5}$]<br>$N_X$ |
|---|---|---|---|---|
| 1H0419-577 | | | | |
| 112600401 | 21.15 ± 0.17 | 2.00 ± 0.09 | 4.80 ± 0.65 | 3.08 ± 1.52 |
| 148000201 | 22.48 ± 0.10 | 2.10 ± 0.08 | 2.08 ± 0.27 | 4.38 ± 0.79 |
| 148000301 | 21.80 ± 0.15 | 1.85 ± 0.35 | 1.99 ± 0.21 | 2.28 ± 0.90 |
| 148000401 | 21.36 ± 0.09 | 0.10 ± 0.16 | 2.90 ± 0.06 | 1.44 ± 0.77 |
| 148000501 | 21.52 ± 0.07 | 2.14 ± 0.03 | 1.25 ± 0.74 | 3.27 ± 2.08 |
| 148000601 | 22.07 ± 0.05 | 2.18 ± 0.02 | 2.12 ± 1.05 | 3.38 ± 0.98 |
| 604720301 | 20.93± 0.08 | 2.00 ± 0.04 | 2.57 ± 0.19 | 1.38 ± 0.45 |
| 604720401 | 21.34 ± 0.05 | 2.23 ± 0.03 | 2.29 ± 0.27 | 1.33 ± 0.58 |
| ESO198-G24 | | | | |
| 67190101 | 19.00* | 3.00* | 3.46 ± 0.56 | 1.63 ± 0.51 |
| 112910101 | 19.00* | 3.00* | 2.68 ± 1.57 | 4.24 ± 1.03 |
| 305370101 | 19.00* | 3.00* | 0.22 ± 0.14 | 10.26 ± 7.22 |
| HB890405-123 | | | | |
| 202210301 | 20.51 ± 0.41 | 1.89 ± 0.54 | 1.10 ± 0.17 | 0.76 ± 0.25 |
| 202210401 | 20.38 ± 0.48 | 1.70 ± 1.29 | 0.86 ± 0.09 | 0.71 ± 0.27 |
| HE1029-1401 | | | | |
| 110950101 | 21.22 ± 0.79 | 3.00 ± 0.69 | 2.03 ± 0.71 | 3.60 ± 1.27 |
| 203770101 | 20.61 ± 0.22 | 2.43 ± 0.20 | 5.06 ± 0.44 | 4.01 ± 0.64 |
| IRASF12397+3333 | | | | |
| 202180201 | 22.17 ± 0.01 | 2.16 ± 0.01 | 1.20 ± 0.70 | 1.17 ± 0.25 |
| 202180301 | 22.27 ± 0.04 | 2.19 ± 0.01 | 1.73 ± 0.30 | 1.22 ± 0.71 |
| LBQS1228+1116 | | | | |
| 306630101 | 21.36 ± 0.16 | 2.13 ± 0.09 | 0.30 ± 0.07 | 0.30 ± 0.13 |
| 306630201 | 21.47 ± 0.12 | 2.17 ± 0.06 | 0.33 ± 0.07 | 0.37 ± 0.12 |
| MRK279 | | | | |
| 302480401 | 20.88 ± 0.07 | 1.96 ± 0.04 | 7.68 ± 0.51 | 8.52 ± 0.73 |
| 302480501 | 20.70 ± 0.07 | 1.56 ± 0.18 | 6.91 ± 0.31 | 7.79 ± 0.66 |
| 302480601 | 20.53 ± 0.11 | 1.00 ± 0.85 | 7.19 ± 0.60 | 9.52 ± 0.96 |
| MRK335 | | | | |
| 510010701 | 22.50 ± 0.05 | 1.89 ± 0.06 | 3.31 ± 1.03 | 6.22 ± 2.63 |
| 600540501 | 22.01 ± 0.07 | 2.12 ± 0.08 | 2.27 ± 0.19 | 4.92 ± 0.36 |
| 600540601 | 22.22 ± 0.05 | 2.14 ± 0.01 | 1.70 ± 0.06 | 3.50 ± 0.20 |
| MRK509 | | | | |
| 130720101 | 21.56 ± 0.05 | 2.17 ± 0.02 | 6.87 ± 0.57 | 9.68 ± 1.04 |
| 601390201 | 21.32 ± 0.03 | 2.18 ± 0.02 | 8.99 ± 0.88 | 10.73 ± 0.90 |
| 601390301 | 21.36 ± 0.03 | 2.16 ± 0.01 | 9.68 ± 1.53 | 13.44 ± 1.01 |
| 601390401 | 21.45 ± 0.02 | 1.59 ± 0.01 | 9.25 ± 0.99 | 14.95 ± 0.92 |
| 601390501 | 21.37 ± 0.03 | 2.15 ± 0.02 | 8.05 ± 1.46 | 11.32 ± 1.02 |
| 601390601 | 21.38 ± 0.03 | 2.15 ± 0.02 | 7.86 ± 0.66 | 12.63 ± 0.94 |
| 601390701 | 21.39 ± 0.03 | 2.19 ± 0.01 | 11.58 ± 2.87 | 12.56 ± 1.16 |
| 601390801 | 21.37 ± 0.03 | 2.18 ± 0.01 | 9.85 ± 1.38 | 13.97 ± 0.95 |
| 601390901 | 21.37 ± 0.03 | 2.22 ± 0.01 | 8.47 ± 1.17 | 13.17 ± 1.04 |
| 601391001 | 21.35 ± 0.03 | 2.17 ± 0.01 | 11.48 ± 1.54 | 12.91 ± 0.93 |
| 601391101 | 21.39 ± 0.02 | 2.18 ± 0.01 | 9.62 ± 1.12 | 13.64 ± 0.96 |
| MRK590 | | | | |
| 109130301 | 20.30* | 3.00* | 0.88 ± 0.63 | 4.82 ± 1.24 |
| 201020201 | 19.00* | 3.00* | 0.01 ± 0.29 | 4.14 ± 5.09 |
| NGC4593 | | | | |
| 59830101 | 21.74 ± 0.04 | 2.42 ± 0.06 | 12.25 ± 0.11 | 18.09 ± 0.75 |
| 109970101 | 21.86 ± 0.04 | 2.18 ± 0.02 | 5.12 ± 1.28 | 17.09 ± 2.24 |
| 740920201 | 21.64 ± 0.36 | 2.71 ± 0.37 | 6.28 ± 0.22 | 9.56 ± 1.13 |
| 740920301 | 21.16 ± 0.09 | 0.25 ± 0.14 | 2.95 ± 0.15 | 11.45 ± 1.03 |
| 740920401 | 21.45 ± 0.16 | 2.07 ± 0.06 | 3.38 ± 0.13 | 11.71 ± 0.88 |
| 740920501 | 21.80 ± 0.04 | 2.18 ± 0.02 | 4.59 ± 1.77 | 14.74 ± 2.10 |
| 740920601 | 21.86 ± 0.13 | 2.70 ± 0.03 | 5.79 ± 2.17 | 12.12 ± 1.28 |
| NGC7469 | | | | |
| 112170101 | 21.75 ± 0.13 | 2.52 ± 0.10 | 8.75 ± 0.52 | 13.05 ± 1.35 |
| 112170301 | 21.74 ± 0.03 | 1.95 ± 0.07 | 9.91 ± 0.20 | 10.79 ± 1.12 |
| 207090101 | 21.65 ± 0.05 | 2.26 ± 0.01 | 3.52 ± 0.46 | 12.76 ± 1.12 |
| 207090201 | 21.00 ± 0.10 | 2.14 ± 0.04 | 9.08 ± 0.41 | 8.91 ± 0.47 |
| PG0804+761 | | | | |
| 605110101 | 20.12 ± 3.24 | 3.00 ± 3.00 | 0.95 ± 0.17 | 3.28 ± 1.02 |
| 605110201 | 21.12 ± 0.51 | 3.00± 0.45 | 0.52 ± 0.15 | 5.83 ± 1.47 |
| PG0844+349 | | | | |

Table D.3: continued.

| Source<br>ObsID | log $N_H^{CL}$ | log $\xi_{CL}$ | [× $10^{-3}$]<br>$N_{hc}$ | [× $10^{-5}$]<br>$N_X$ |
|---|---|---|---|---|
| 103660201 | 21.01 ± 0.16 | 1.92 ± 0.10 | 2.03 ± 0.30 | 1.06 ± 0.57 |
| 554710101 | 21.87 ± 0.19 | 1.99 ± 0.15 | 0.09 ± 0.02 | 2.05 ± 0.76 |
| PG1116+215 | | | | |
| 201940101 | 20.64 ± 0.15 | 1.69 ± 0.34 | 0.92 ± 0.17 | 0.76 ± 0.21 |
| 201940201 | 19.00* | 3.00* | 1.48 ± 0.31 | 1.53 ± 0.95 |
| 554380101 | 20.53 ± 0.23 | 1.70 ± 0.47 | 0.85 ± 0.73 | 0.93 ± 0.38 |
| 554380201 | 20.75 ± 0.16 | 1.71 ± 0.31 | 0.56 ± 0.26 | 1.13 ± 0.41 |
| 554380301 | 20.76 ± 0.16 | 1.71 ± 0.36 | 0.36 ± 0.06 | 1.44 ± 0.36 |
| PG1351+640 | | | | |
| 205390301 | 21.70 ± 0.05 | 1.00 ± 0.16 | 0.23 ± 0.05 | 0.12 ± 0.12 |
| 556230201 | 22.38 ± 0.23 | 2.16 ± 0.01 | 0.02 ± 0.06 | 0.34 ± 0.22 |
| PG1402+261 | | | | |
| 400200101 | 21.09 ± 0.21 | 2.06 ± 0.09 | 0.62 ± 0.15 | 0.61 ± 0.31 |
| 400200201 | 20.74 ± 0.32 | 1.10 ± 1.70 | 0.47 ± 0.17 | 1.05 ± 0.67 |
| PG1440+356 | | | | |
| 5010101 | 21.10 ± 0.14 | 2.04 ± 0.07 | 0.71 ± 0.65 | 1.14 ± 1.63 |
| 5010201 | 20.31 ± 0.40 | 1.00 ± 2.49 | 0.95 ± 0.34 | - |
| 5010301 | 20.93 ± 0.17 | 1.83 ± 0.30 | 0.76 ± 0.26 | 0.26 ± 0.32 |
| 107660201 | 20.88 ± 0.08 | 1.00 ± 0.74 | 1.43 ± 0.24 | 0.67 ± 0.49 |
| Q0056-363 | | | | |
| 102040701 | 20.86 ± 0.36 | 1.90 ± 0.31 | 1.03 ± 0.30 | 1.54 ± 0.56 |
| 205680101 | 20.90 ± 0.10 | 1.49 ± 0.20 | 0.29 ± 0.07 | 1.09 ± 0.24 |
| 401930101 | 21.06 ± 0.14 | 1.93 ± 0.01 | 0.88 ± 0.21 | 0.47 ± 0.22 |
| RE1034+396 | | | | |
| 109070101 | 20.49 ± 0.46 | 1.69 ± 0.97 | 0.49 ± 0.10 | 1.08 ± 0.62 |
| 506440101 | 20.49 ± 0.14 | 1.49 ± 0.42 | 0.56 ± 0.06 | 0.44 ± 0.21 |
| 561580201 | 20.60 ± 0.13 | 1.29 ± 0.60 | 0.38 ± 0.04 | 0.42 ± 0.29 |
| 655310101 | 21.56 ± 0.51 | 2.90 ± 0.20 | 0.46 ± 0.06 | - |
| 655310201 | 22.24 ± 0.28 | 3.00 ± 0.19 | 0.30 ± 0.07 | - |
| 675440101 | 20.95 ± 0.07 | 0.12 ± 0.13 | 0.32 ± 0.06 | - |
| 675440201 | 22.37 ± 0.96 | 3.23 ± 0.20 | 0.31 ± 0.09 | 0.85 ± 0.48 |
| 675440301 | 21.92 ± 0.36 | 3.00 ± 0.23 | 0.34 ± 0.06 | - |
| UGC 3973 | | | | |
| 103862101 | 21.95 ± 0.07 | 1.68 ± 0.12 | 1.64 ± 0.83 | 8.18 ± 2.25 |
| 400070201 | 22.26 ± 0.07 | 2.83 ± 0.05 | 1.85 ± 0.79 | 10.55 ± 2.95 |
| 400070301 | 21.91 ± 0.03 | 2.16 ± 0.01 | 0.34 ± 0.62 | 25.67 ± 46.03 |
| 400070401 | 22.17 ± 0.09 | 2.83 ± 0.08 | 1.21 ± 0.43 | 15.22 ± 3.81 |
| 502091001 | 21.48 ± 0.05 | 1.00 ± 0.32 | 2.79 ± 0.29 | 10.69 ± 0.62 |

**Note:** '*' denotes frozen values and '-' for last column denotes negligible reflection component.